\documentclass{article}%
\usepackage{amsmath}
\usepackage{amsfonts}
\usepackage{amssymb}
\usepackage{graphicx}
\usepackage{epstopdf}
\usepackage[colorlinks]{hyperref}
\usepackage{amsmath,hyperref}
\usepackage{cite}
\usepackage{hyperref}
\usepackage{url}
\newcommand{\bpartial}{\mathop{\partial\kern -4pt\raisebox{.8pt}{$|$}}}

\newcommand{\bra}{\mathopen{[\kern-1.6pt[}}
\newcommand{\ket}{\mathclose{]\kern-1.5pt]}}
\newcommand{\bbra}{\mathopen{[\kern-2.2pt[\kern-2.3pt[}}
\newcommand{\bket}{\mathclose{]\kern-2.1pt]\kern-2.3pt]}}

\makeindex
\begin{document}
\title{\bf  Integrable sigma models with complex and generalized complex structures }
\author { A. Rezaei-Aghdam \hspace{-3mm}{ \footnote{e-mail: rezaei-a@azaruniv.ac.ir}\hspace{2mm}
{\small and }
A. Taghavi \hspace{-3mm}{ \footnote{ e-mail: amanetaghavi@gmail.com}}\hspace{2mm}} \\
{\small{\em Department of Physics, Faculty of science, Azarbaijan  Shahid Madani University }}\\
{\small{\em  53714-161, Tabriz, Iran  }}}

\maketitle
\begin{abstract}
Using the general method presented by Mohammedi \cite{NM} for the integrability of a sigma model on a manifold, we investigate the conditions for having an integrable deformation of the general sigma model on a manifold with a complex structure. On a Lie group, these conditions are satisfied by using the zeros of the Nijenhuis tensor. We then extend this formalism to a manifold, especially a Lie group, with a generalized complex structure and in this manner we present new integrable sigma models. The we demonstrate that, for the examples of integrable sigma models with generalized complex structures on the Lie groups $\mathbf{A_{4,8}}$ and $\mathbf{A_{4,10}}$, under special conditions, the perturbed terms of the actions are identical to the WZ terms.
\end{abstract}
\newpage

\section{\bf Introduction}
Two-dimensional integrable $\sigma$ models and their deformations have always received considerable attention, dating back to the early days of their study \cite{KP, HE}. The integrable deformation of the principal chiral model on $SU(2)$ was first presented in \cite{BP, CI, VA} (for a general Lie group, refer to \cite{NJ}). The Yang-Baxter (or $\eta$) deformation of the chiral model, as a generalization of \cite{CI, VA}, was introduced by Klim$\text{\v{c}}\text{'{\i}}$k \cite{CK, CKL, CKL1}. Furthermore, the $\lambda$ deformation, proposed in \cite{KS}, is a generalization of \cite{BP}. The Yang-Baxter integrable deformation \cite{CK} is based on R-operators that satisfy the (modified) classical Yang-Baxter equation ((m)CYBE) \cite{EP, EP1}. The integrable sigma model on a Lie group with a complex structure was also studied in \cite{BK} as a special case of the Yang-Baxter sigma model \cite{CK}. Recently, Mohammedi proposed a deformation of $WZW$ models using two invertible linear operators \cite{NMO}, and this model consists of the Yang-Baxter deformation as a special case.

Here, we utilize the general method presented by Mohammedi \cite{NM} to construct an integrable sigma model on a general manifold with a complex structure. We establish conditions for integrability and demonstrate that, when the target space is a Lie group, these conditions are automatically satisfied through the use of the Nijenhuis tensor's zeros. Subsequently, we employ this method to develop an integrable sigma model on a manifold, with a particular focus on Lie groups, equipped with a generalized complex structure \cite{HI, GU}. The structure of this paper is as follows:

Section 2: In this section, we review the general method presented by Mohammedi in \cite{NM} for the integrability of a sigma model on a manifold. Section 3: We present an integrable sigma model on a manifold with a complex structure. Subsequently, we apply the method from Section 2 to investigate the integrability of the model and provide two examples. Section 4: Here, we construct an integrable sigma model on a general Lie group with a complex structure. In this case, the integrability conditions are automatically satisfied as a consequence of the Nijenhuis condition. We also provide examples on $\mathbf{A_{4,8}}$ (Heisenberg) and $\mathbf{G_{6,23}}$ Lie groups. Section 5: As a generalization, we present an integrable sigma model on a manifold and a Lie group with generalized complex structures. Additionally, at the end of this section, we demonstrate that for two examples on $\mathbf{A_{4,8}}$ and $\mathbf{A_{4,10}}$ (Nappi-Witten) Lie groups, under special conditions, the perturbed terms of the actions coincide with the WZ terms. Section 6: In this section, we perturb the $WZW$ model using a generalized complex structure on their Lie group. To facilitate comparison with Mohammedi's work \cite{NMO}, we present the generalized complex structure formulas on a metric Lie algebra in terms of operator formalism. Section 7 include some Concluding remarks.
\section{\bf Review of the zero curvature representation and integrability conditions for
 non-linear sigma models}\label{sec2}
In this section, we begin by presenting the notation and reviewing key aspects of the general formalism introduced by Mohammedi \cite{NM} for ensuring the integrability of a non-linear sigma model on a manifold. To illustrate these concepts, let us consider the following two-dimensional sigma model action:\footnote{As \cite{NMO} and \cite{NM} instead using of the world sheet coordinates $\tau$ and $\sigma$, here we will use the complex coordinates ($z=\sigma+i\tau,\bar{z}=\sigma-i\tau$) with $\partial=\frac{\partial}{\partial z}$ and $\bar{\partial}=\frac{{\partial}}{\partial\bar{z}}$. We will also use the convention that the Levi-Civita tensor $\epsilon^{z\bar{z}}=1$; so we will not have $i$ in front of $B$ field in (\ref{eqn:sm}).}
\begin{equation}\label{eqn:sm}
S=\int_{\Sigma}\hspace{2mm}dzd\bar{z}(G_{\mu\nu}(x)+B_{\mu\nu}(x))\hspace{0mm}\partial\hspace{0mm}x^{\mu}\bar{\partial}x^{\nu},
\end{equation}
where $x^{\mu}(z,\bar{z})$ $(\mu=1,2,...,d)$ are coordinates of $d$ dimensional manifold $M$, with $G_{\mu\nu}$ and $B_{\mu\nu}$ as invertible metric and anti-symmetric tensor fields on it and $(z,\bar{z})$ are coordinates of the world sheet $\Sigma$. The equations of motion of this model have the following form :
\begin{equation}\label{eom}
  \bar{\partial} \partial\hspace{0cm}X^{\lambda}+\Omega^{\lambda}\hspace{0cm}_{\mu\nu}\partial\hspace{0cm}x^{\mu}\bar{\partial}x^{\nu}=0,
\end{equation}
where
\begin{equation}
 \Omega^{\lambda}\hspace{0cm}_{\mu\nu}=\Gamma^{\lambda}\hspace{0cm}_{\mu\nu}-H^{\lambda}\hspace{0cm}_{\mu\nu},
\end{equation}
such that $\Gamma^{\lambda}\hspace{0cm}_{\mu\nu}$ are
Christoffel coefficients and the components of torsion $H^{\lambda}\hspace{0cm}_{\mu\nu}$ are given by

\begin{equation}
H^{\lambda}\hspace{0cm}_{\mu\nu}=\frac{1}{2}G^{\lambda\eta}(\partial_{\eta}B_{\mu\nu}+\partial_{\nu}B_{\eta\mu}+\partial_{\mu}B_{\nu\eta}).
\end{equation}

According to \cite{NM}, the sigma model (\ref{eqn:sm}) is integrable if one can construct the following linear system, whose consistency
conditions (a zero curvature representation) are equivalence to
the equations of motion\footnote{In general the Lax equation can be written as
\begin{align}
[\partial_{0}+\mathcal{A}_{0} (x,\lambda)]\psi=0,
\notag
\end{align}
\begin{equation}
[\partial_{1}+\mathcal{A}_{1} (x,\lambda)]\psi=0,
\end{equation}
where the matrices $\mathcal{A}_{0}$ and $\mathcal{A}_{1}$ depend on the fields $x$ and spectral parameter $\lambda$. This parameter is very useful in obtaining conserved quantities. Furthermore the field $\psi$ can be either a column vector or a matrix of the same dimension as $\mathcal{A}_{0}$ and $\mathcal{A}_{1}$. The consistency condition (the zero curvature condition) of this linear system is $(\partial_{0}\mathcal{A}_{1}-\partial_{1}\mathcal{A}_{0}+[\mathcal{A}_{0},\mathcal{A}_{1}])\psi=0$. The conserved quantities are constructed using the monodromy matrix
\begin{align}
\mathcal{M}(\lambda,\tau)=P exp(-\int^{2\pi}_{0} \mathcal{A}_{1}(\lambda,\sigma,\tau) d\sigma ),
\end{align}
where $P$ is the path-ordered exponential. The traces of powers of the monodromy matrix, $Tr[\mathcal{M}^{n}(\lambda,\tau)]$ are involution with respect to Poisson brackets: $\{Tr[\mathcal{M}^{m}(\lambda_{1},\tau)],Tr[\mathcal{M}^{n}(\lambda_{2},\tau)]\}=0$ and are independent of the time $\tau$. Expanding $Tr[\mathcal{M}^{n}(\lambda,\tau)]$ in powers of $\lambda$ give an infinite set of conserved charges \cite{HE,FTB}.
} (\ref{eom})
\begin{align}
[\partial+\partial\hspace{0cm}x^{\mu}.\alpha_{\mu}(x,\lambda)]\psi=0,
\notag
\end{align}
\begin{equation}\label{lax}
[\bar{\partial}+\bar{\partial}x^{\nu}.\beta_{\nu}(x,\lambda)]\psi=0,
\end{equation}
where the matrices $\alpha_{\mu}(x,\lambda)$ and $\beta_{\mu}(x,\lambda)$ are functions of coordinates $x^{\mu}$ and spectral parameter $\lambda$.
The compatibility condition of this linear system yields the
equations of motion if the matrices $\alpha_{\mu}(x,\lambda)$
and $\beta_{\mu}(x,\lambda)$ satisfies the following relation \cite{NM}:
\begin{equation}\label{me1}
 \partial\hspace{0cm}_{\mu}\beta_{\nu}-\partial\hspace{0cm}_{\nu}\alpha_{\mu}+[\alpha_{\mu},\beta_{\nu}]=\Omega^{\lambda}\hspace{0cm}_{\mu\nu}\mu_{\lambda},
\end{equation}
such that with $\beta_{\mu}-\alpha_{\mu}=\mu_{\mu}$ the equation (\ref{me1}) can be rewritten as
\begin{equation}\label{me}
 F_{\mu\nu}=-(\nabla_{\mu}\mu_{\nu}-\Omega^{\lambda}\hspace{0cm}_{\mu\nu}\mu_{\lambda})
 ,
\end{equation}
where the field strength $F_{\mu\nu}$ and covariant derivative with respect to the matrices $\alpha_{\mu}$ are given as follows:
\begin{equation}
 F_{\mu\nu}=\partial_{\mu}\alpha_{\nu}-\partial_{\nu}\alpha_{\mu}+[\alpha_{\mu},\alpha_{\nu}]\hspace{1cm},\hspace{1cm}
 \nabla_{\mu}V=\partial_{\mu}V+[\alpha_{\mu},V].
\end{equation}
Note that by splitting the equation (\ref{me}) into its symmetric and anti-symmetric components, we can express it as the following set of relations \cite{NM}{\footnote{Note that the equation (\ref{eqn:sp}) is a gauged version of a matrix-valued Killing equation. Indeed, if $[\alpha_{\mu},\mu_{\nu}]+[\alpha_{\nu},\mu_{\mu}]=0$, then this equation is simplify as Killing one $\partial_{\mu}\mu_{\nu}+\partial_{\nu}\mu_{\mu}-2\Gamma^{\lambda}_{\mu\nu}\mu_{\lambda}=0$ \cite{NM}.}}:
\begin{equation}\label{eqn:sp}
 0=\nabla_{\mu}\mu_{\nu}+\nabla_{\nu}\mu_{\mu}-2\Gamma^{\lambda}\hspace{0cm}_{\mu\nu}\mu_{\lambda},
\end{equation}
\begin{equation}\label{eqn:asp}
 F_{\mu\nu}=-\frac{1}{2}(\nabla_{\mu}\mu_{\nu}-\nabla_{\nu}\mu_{\mu})-H^{\lambda}\hspace{0cm}_{\mu\nu}\mu_{\lambda}.
\end{equation}
 Thus, the integrability condition of the sigma model (\ref{eqn:sm}) is equivalent to finding matrices $\alpha_{\mu}$ and $\mu_{\mu}$ that satisfy either the single relation (\ref{me}) or the pair of relations (\ref{eqn:sp}) and (\ref{eqn:asp}).
\section{\bf  Integrable sigma model with complex structure on manifold}\vspace{0.5cm}\label{sec3}
Here we will attempt to construct an integrable sigma model on a manifold with a complex structure. Let $M$ be a differential manifold; the pair $(M, J)$ is called an almost complex manifold if there exists a tensor field $J$ of $(1,1)$ type such that at each point $p$ of $M$, $J^2_p = -1$. The tensor field $J$ is referred to as the almost complex structure. An almost complex structure $J$ on a manifold $M$ is integrable if and only if its Nijenhuis tensor vanishes \cite{NA}
\begin{equation}\label{nij0}
N(X,Y)=0\hspace{1cm},\hspace{1cm}\forall X,Y \in \chi(M),
\end{equation}
where $\chi(M)$ is the set of vector fields on $M$ and the Nijenhuis tensor $N: \chi(M)\otimes \chi(M) \longrightarrow \chi(M)$ is given by

\begin{equation}\label{nij1}
N(X,Y)=J^2[X,Y]-J[JX,Y]-J[X,JY]+[JX,JY].
\end{equation}
In the coordinate basis, represented by the basis vectors $e_{\mu}$ and dual vectors (forms) $dx^{\mu}$ on the manifold $M$, the almost complex structure and the Nijenhuis tensor are expressed as $J=J^{\mu}\hspace{0cm}_{\nu}e_{\mu}\otimes dx^{\nu}$ and $N=N^{\lambda}\hspace{0cm}_{\mu\nu}e_{\lambda}\otimes dx^{\mu}\otimes dx^{\nu}$
respectively and the integrability condition (\ref{nij0}) can be rewritten as follows:
\begin{equation}\label{nj}
N^{\mu}\hspace{0cm}_{\nu k}=J^{\lambda}\hspace{0cm}_{\nu}\partial_{\lambda}J^{\mu}\hspace{0cm}_{k}-J^{\lambda}\hspace{0cm}_{k}\partial_{\lambda}J^{\mu}\hspace{0cm}_{\nu}-J^{\mu}\hspace{0cm}_{\lambda}\partial_{\nu}J^{\lambda}\hspace{0cm}_{k}+
J^{\mu}\hspace{0cm}_{\lambda}\partial_{k}J^{\lambda}\hspace{0cm}_{\nu}=0,
\end{equation}
also the relation $J^{2}=-1$ can be given as
\begin{equation}\label{jt}
J^{\mu}\hspace{0cm}_{\lambda}J^{\lambda}\hspace{0cm}_{\nu}=-\delta^{\mu}\hspace{0cm}_{\nu}.
\end{equation}
Now on the manifold $M$ with coordinates $x^{\mu}$, metric $g_{\mu\nu}$ and a complex structure $J^{\mu}\hspace{0cm}_{\nu}$, we propose the following deformed sigma model with complex structure:
\begin{equation}\label{csm}
S=\int\hspace{2mm}dzd\bar{z}(g_{\mu\nu}+k
g_{\mu\lambda}J^{\lambda}\hspace{0mm}_{\nu}
)\partial\hspace{0mm}x^{\mu}\bar{\partial}x^{\nu},
\end{equation}
where $k$ is a constant parameter. In the following, we will demonstrate the integrability of this model under certain additional conditions on the complex structure $J^\mu\hspace{0cm}_{\nu}$. To achieve this, we will employ the method described in the previous section. By comparing this action with (\ref{eqn:sm}), we can observe that
\begin{equation}
G_{\mu\nu}=g_{\mu\nu}+\dfrac{k}{2}(g_{\mu\lambda}J^{\lambda}\hspace{0cm}_{\nu}+g_{\nu\lambda}J^{\lambda}\hspace{0cm}_{\mu}),
\end{equation}
\begin{equation}
B_{\mu\nu}=\dfrac{k}{2}(g_{\mu\lambda}J^{\lambda}\hspace{0cm}_{\nu}-g_{\nu\lambda}J^{\lambda}\hspace{0cm}_{\mu}).
\end{equation}
For invertibility of the metric $G_{\mu\nu}$ we assume the following form for $G^{\mu\nu}$:
\begin{equation}
G^{\mu\nu}=a g^{\mu\nu}+b(g^{\mu\lambda}J^{\nu}\hspace{0cm}_{\lambda}+g^{\nu\lambda}J^{\mu}\hspace{0cm}_{\lambda}),
\end{equation}
where $a$ and $b$ are constant parameters. Now using $G^{\mu\lambda}G_{\lambda\nu}=\delta^{\mu}\hspace{0cm}_{\nu}$
, $J^{\mu}\hspace{0cm}_{\lambda}J^{\lambda}\hspace{0cm}_{\nu}=-\delta^{\mu}\hspace{0cm}_{\nu}$ and applying Hermitian (metric compatible) complex structure
\begin{equation}\label{hcsc}
 J^{\mu}\hspace{0cm}_{\nu}=-g^{\mu\lambda}J^{\rho}\hspace{0cm}_{\lambda}g_{\rho\nu}
 \end{equation}
with $a = 1$ and the arbitrary $b$, then we have $G_{\mu\nu}=g_{\mu\nu}$.
For requiring integrability of the sigma model (\ref{csm}) one can consider matrices $\alpha_{\mu}$ and $\mu_{\mu}$ as
\begin{equation}
\alpha_{\mu}=\lambda_{1}c_{\lambda}J^{\lambda}\hspace{0.0cm}_{\mu}
  \hspace{0.5cm},\hspace{0.5cm}
\mu_{\mu}=\lambda_{2}d_{\lambda}J^{\lambda}\hspace{0.0cm}_{\mu},
\end{equation}
or using veilbein formalism ($e_{\mu}=\widehat{e}_{\beta}e^{\beta}\hspace{0.0cm}_{\mu}, dx^{\mu}=e_{\alpha}\hspace{0.0cm}^{\mu}\widehat{\theta}^{\alpha}, e^{\alpha}\hspace{0.0cm}_{\mu}e_{\beta}\hspace{0.0cm}^{\mu}=\delta^{\alpha}\hspace{0.0cm}_{\beta}, e^{\alpha}\hspace{0.0cm}_{\mu}e_{\alpha}\hspace{0.0cm}^{\nu}=\delta_{\mu}\hspace{0.0cm}^{\nu}$) \cite{NA} one can rewrite matrices $\alpha_{\mu}$ and $\mu_{\mu}$ as follows
\begin{equation}\label{am1}
\alpha_{\mu}=\lambda_{1}c_{\alpha}J^{\alpha}\hspace{0.0cm}_{\beta}e^{\beta}\hspace{0.0cm}_{\mu}
  \hspace{0.5cm},\hspace{0.5cm}
\mu_{\mu}=\lambda_{2}d_{\alpha}J^{\alpha}\hspace{0.0cm}_{\beta}e^{\beta}\hspace{0.0cm}_{\mu},
\end{equation}
with
\begin{equation}
J^{\mu}\hspace{0.0cm}_{\nu}=e_{\alpha}\hspace{0.0cm}^{\mu}J^{\alpha}\hspace{0.0cm}_{\beta}e^{\beta}\hspace{0.0cm}_{\nu}.
\end{equation}
In our formulation, we assume that $c_{\alpha}$ and $d_{\alpha}$ are square matrices with constant elements, and $\lambda_{1}$ and $\lambda_{2}$ are constant parameters. Additionally, $J^{\alpha}\hspace{0.0cm}_{{\beta}}$ represents constant algebraic complex structure coefficients. By this assumption for the matrices $\alpha_{\mu}$ and $\mu_{\mu}$
 let us investigate some examples.
\vspace{0.5cm}

{\bf 3.1 Examples}\vspace{0.5cm}

 Here we consider two examples for the integrable sigma model (\ref{csm}).
 \vspace{0.3cm}

\textbf{a}) As the first example, we consider four dimensional manifold as a product of two 2d sausage model\cite{FV} as follows \footnote{Note that the integrability of sausage model is conjectured in \cite{FV} and then proven in \cite{LS}. When the parameter $\chi$ is zero, the metric correspond to round $S^2$ of radius $\sqrt{h}$. For the real value of  $\chi$ it is proven that the sausage model is a Yang-Baxter sigma model \cite{HB}. For  $\chi^2=-1$ case the 2d sausage metric reduce to the metric of $\frac{SO(1,2)}{SO(2)}$ gauged $WZW$ model \cite{HN}.}:
\begin{equation}\label{sausag}
ds^2=\frac{1}{(1-r^2)(1+\chi^2r^2)}dr^2+\frac{1-r^2}{1+\chi^2r^2}d\varphi^2+\frac{f(\varphi)}{(1+\rho^2)(1-\chi^2\rho^2)}d\rho^2+f(\varphi)\frac{1+\rho^2}{1-\chi^2\rho^2}dt^2,
\end{equation}
 Here we will try to deform this model with Hermitian complex structure $J$ and show that the deformed model is integrable. The tensor field $J$ that satisfies the integrable complex structure conditions (\ref{nj}), (\ref{jt}) and the Hermitian complex structure (\ref{hcsc}) can be obtained as follows
\begin{equation}
(J^{\mu}\hspace{0.0cm}_{\nu})=\left(
\begin{array}{cccc}
  0 & 1-r^2 &0&0\\
 \frac{-1}{1-r^2} &0&0&0\\
  0 & 0 &0&-(1+\rho^2)\\
 0 &0&\frac{1}{1+\rho^2}&0\\
  \end{array}\right),
\end{equation}
where the veilbeins and an algebraic components of $J^{\alpha}\hspace{0.0cm}_{\beta}$ are given as
\begin{equation}
(e^{\alpha}\hspace{0.0cm}_{\mu})=\left(
\begin{array}{cccc}
  \sqrt{\frac{1}{(1-r^2)(1+\chi^2r^2)}}&0&0&0 \\
 0 &\sqrt{\frac{1-r^2}{1+\chi^2r^2}}&0&0\\
 0&0&\sqrt{\frac{f(\varphi)}{(1+\rho^2)(1-\chi^2\rho^2)}}&0 \\
 0 &0&0&\sqrt{f(\varphi)\frac{1+\rho^2}{1-\chi^2\rho^2}}\\
  \end{array}\right)\hspace{0.5cm},\hspace{0.5cm}(J^{\alpha}\hspace{0.0cm}_{\beta})=\left(
\begin{array}{cccc}
  0 &1 &0&0\\
 -1 &0&0&0\\
  0 &0 &0&-1\\
 0&0&1&0\\
  \end{array}\right).
\end{equation}
The integrable sigma model ($\ref{csm}$) can be expressed as
\begin{equation*}
S=\int dzd\overline{z}[\frac{1}{(1-r^2)(1+\chi^2r^2)}\partial r\overline{\partial}r+\frac{1-r^2}{1+\chi^2r^2}\partial \varphi\overline{\partial }\varphi+\frac{f(\varphi)}{(1+\rho^2)(1-\chi^2\rho^2)}\partial \rho\overline{\partial}\rho+\frac{f(\varphi)(1+\rho^2)}{1-\chi^2\rho^2}\partial t \overline{\partial }t
\end{equation*}
\begin{equation}\label{sa}
-\frac{kf(\varphi)}{(1-\chi^2\rho^2)}(\partial \rho\overline{\partial }t-\partial t\overline{\partial }\rho)].
\end{equation}
Note that in this and all subsequent examples in the paper, we have omitted the surface terms \footnote{For this action, the term $\frac{k}{(1+\chi^2r^2)}(\partial r\overline{\partial }\varphi-\partial\varphi\overline{\partial }r)$ is a surface term and can be omitted.} in the actions.
One can investigate the integrability condition (\ref{me}) for this model. By setting $\chi^2=-1$,  the Christoffel and torsion components of this model are given as follows
\begin{equation*}
\Gamma^{1}\hspace{0cm}_{11}=-\frac{2r}{(-1+r^2)},\hspace{0.5cm}\Gamma^{2}\hspace{0cm}_{33}=-\frac{f'(\varphi)}{2(1+\rho^2)^2},\hspace{0.5cm}\Gamma^{2}\hspace{0cm}_{44}=-\frac{f'(\varphi)}{2} ,
\end{equation*}
\begin{equation*}
\Gamma^{3}\hspace{0cm}_{23}=\frac{f'(\varphi)}{2f(\varphi)} ,\hspace{0.5cm}\Gamma^{3}\hspace{0cm}_{33}=-\frac{2\rho}{(1+\rho^2)} ,\hspace{0.5cm}\Gamma^{4}\hspace{0cm}_{24}=\frac{f'(\varphi)}{2f(\varphi)},
\end{equation*}
\begin{equation}
H^{2}\hspace{0cm}_{34}=-\frac{kf'(\varphi)}{2(1+\rho^2)},\hspace{0.5cm}  H^{3}\hspace{0cm}_{24}=\frac{k(1+\rho^2)f'(\varphi)}{2f(\varphi)}
,\hspace{0.5cm} H^{4}\hspace{0cm}_{32}=\frac{kf'(\varphi)}{2f(\varphi)(1+\rho^2)},
 \end{equation}
then by applying the matrices $\alpha_{\mu} $ and $\mu_{\mu} $ (\ref{am1}) and furthermore matrices $c_{3}=c_{4}=d_{1}=d_{3}=d_{4}=0$ and setting matrices $c_{1},c_{2},d_{2}$ as the following form\footnote{The matrix $c_{1},c_{2},d_2$ can be included arbitrary elements as spectral parameter\cite{NM}.}
\begin{equation}
c_{1}=\left(
\begin{array}{cccc}
 m_{1} &0&0&n_{1} \\
 0 &r_{1}&s_{1}&0\\
 0&s_{1}&r_{1}&0\\
 n_{1}&0&0&m_{1}\\
   \end{array}\right),\hspace{0.5cm}c_{2}=\left(
\begin{array}{cccc}
  m_{2} &0&0&n_{2} \\
 0 &r_{2}&s_{2}&0\\
 0&s_{2}&r_{2}&0\\
 n_{2}&0&0&m_{2}\\
  \end{array}\right),\hspace{0.5cm}d_{2}=\left(
\begin{array}{cccc}
  m_{3} &0&0&n_{3} \\
 0 &r_{3}&s_{3}&0\\
 0&s_{3}&r_{3}&0\\
 n_{3}&0&0&m_{3}\\
  \end{array}\right),
\end{equation}
the integrability condition (\ref{me}) satisfies. Note that the $m_i, r_i, s_i, n_i$ are arbitrary (not all simultaneously zero) real constants, and one of them can be considered as spectral parameter.
\vspace{0.5cm}

\textbf{b}) As a second example we consider four-dimensional manifold with the following spherical metric \cite{NR}
\begin{equation}\label{sg}
ds^2=f(r)dr^2+e(r)r^2d\theta^2+e(r)r^2sin^2\theta d\phi^2+h(r)dz^2,
\end{equation}
where $f(r)$, $e(r)$ and $h(r)$ are arbitrary functions of $r$. For this metric the integrable complex structure $J$ which satisfies the relations (\ref{nj}), (\ref{jt}) and (\ref{hcsc})  can be obtained as
\begin{equation}
(J^{\lambda}\hspace{0.0cm}_{\mu})=\left(
\begin{array}{cccc}\label{cj4}
  0 &0&0&-\sqrt{\frac{h(r)}{f(r)}} \\
  0 &0&-sin\theta&0 \\
  0&\frac{1}{sin\theta}&0&0\\
 \sqrt{\frac{f(r)}{h(r)}}&0&0&0\\
 \end{array}\right),
\end{equation}
such that the veilbein and algebraic components of $J^{\alpha}\hspace{0.0cm}_{\beta}$ are as follows:
\begin{equation}
(e^{\alpha}\hspace{0.0cm}_{\mu})=\left(
\begin{array}{cccc}
  \sqrt{f(r)} &0&0&0 \\
 0 &\sqrt{e(r) r^2}&0&0\\
 0&0&\sqrt{e(r) r^2 sin^2\theta}&0\\
 0&0&0&\sqrt{h(r)}\\
   \end{array}\right)\hspace{0.5cm},\hspace{0.5cm}(J^{\alpha}\hspace{0.0cm}_{\beta})=\left(
\begin{array}{cccc}
  0&0&0 &-1 \\
  0&0&-1&0\\
 0&1&0&0\\
 1&0&0&0\\
  \end{array}\right).
\end{equation}
 The action of this model, after omitting surface terms, is given by:
\begin{equation}
S=\int dzd\bar{z}[f(r)\partial r\bar{\partial}r+e(r) r^2(\partial \theta\bar{\partial }\theta+sin^2\theta \partial \phi\bar{\partial }\phi)+D\partial z \bar{\partial }z
+ke(r) r^2sin\theta(\partial \phi\bar{\partial }\theta-\partial \theta\bar{\partial }\phi)].
\end{equation}

Now by applying matrices $\alpha_{\mu}$ and $\mu_{\mu}$ as the relation (\ref{am1}),
the Christoffel components
\begin{equation*}
\Gamma^{1}\hspace{0cm}_{11}=\frac{f'(r)}{2f(r)},\hspace{0.5cm} \Gamma^{1}\hspace{0cm}_{22}=-\frac{r(2e(r)+re'(r))}{2f(r)},\hspace{0.5cm}\Gamma^{1}\hspace{0cm}_{33}=-\frac{rsin^2\theta(2e(r)+re'(r))}{2f(r)},
\notag\end{equation*}
\begin{equation*}
\Gamma^{1}\hspace{0cm}_{44}=-\frac{h'(r)}{2h(r)},\hspace{0.5cm} \Gamma^{2}\hspace{0cm}_{12}=\frac{(2e(r)+re'(r))}{2re(r)},\hspace{0.5cm}\Gamma^{2}\hspace{0cm}_{33}=-cos\theta sin\theta,
\notag\end{equation*}
\begin{equation}
\Gamma^{3}\hspace{0cm}_{13}=\frac{r(2e(r)+re'(r))}{2re(r)},\hspace{0.5cm} \Gamma^{3}\hspace{0cm}_{23}=cot\theta,\hspace{0.5cm}\Gamma^{4}\hspace{0cm}_{14}=\frac{h'(r)}{2h(r)},
\end{equation}
and torsion components of the model this model
\begin{equation}
H^{1}\hspace{0cm}_{23}=-\frac{rsin\theta(2e(r)+re'(r))}{2f(r)},\hspace{0.5cm}  H^{2}\hspace{0cm}_{13}=\frac{sin\theta(2e(r)+re'(r))}{2re(r)}
,\hspace{0.5cm} H^{3}\hspace{0cm}_{21}=\frac{csc\theta(2e(r)+re'(r))}{2re(r)},
 \end{equation}
then by choosing
 $h'(r)=0$ ($h(r)=D$), $c_{2}=c_{3}=d_{2}=d_{3}=d_{4}=0$ and setting matrices $c_{1},c_{4},d_{1}$ as the following form
\begin{equation}
c_{1}=\left(
\begin{array}{cccc}
 m_{1} &0&0&n_{1} \\
 0 &r_{1}&s_{1}&0\\
 0&s_{1}&r_{1}&0\\
 n_{1}&0&0&m_{1}\\
   \end{array}\right),\hspace{0.5cm}c_{4}=\left(
\begin{array}{cccc}
  m_{2} &0&0&n_{2} \\
 0 &r_{2}&s_{2}&0\\
 0&s_{2}&r_{2}&0\\
 n_{2}&0&0&m_{2}\\
  \end{array}\right),\hspace{0.5cm}d_{1}=\left(
\begin{array}{cccc}
  m_{3} &0&0&n_{3} \\
 0 &r_{3}&s_{3}&0\\
 0&s_{3}&r_{3}&0\\
 n_{3}&0&0&m_{3}\\
  \end{array}\right),
\end{equation}

 the metric (\ref{sg}) and complex structure (\ref{cj4}) satisfying the integrability condition (\ref{me}) and making this model integrable. As above the $m_i, r_i, s_i, n_i$ are arbitrary (not all simultaneously zero) real constants, and one of them can be considered as spectral parameter.
\section{\bf Integrable sigma model with complex structure on Lie group }\vspace{0.5cm}\label{sec4}
In the case that $M$ is a Lie group $\mathbf{L}$, using the vielbein formalism
 \begin{equation}\label{vb}
\forall l\in L\hspace{0.5cm},\hspace{0.5cm}(l^{-1}\partial{l})^{\alpha}=e^{\alpha}\hspace{0cm}_{\mu}{\partial}x^{\mu},
\end{equation}
 \begin{equation}\label{gj}
g_{\mu\nu}=e^{\alpha}\hspace{0cm}_{\mu}g_{\alpha\beta}e^{\beta}\hspace{0cm}_{\nu}\hspace{0.5cm},\hspace{0.5cm}J^{\mu}\hspace{0cm}_{\nu}=e_{\alpha}\hspace{0cm}^{\mu}J^{\alpha}\hspace{0cm}_{\beta}e^{\beta}\hspace{0cm}_{\nu},
\end{equation}
the action (\ref{csm}) can be rewritten as follows:\footnote{Note that this model is a special case of the Yang-Baxter sigma model (with $J^2 = -1$) as discussed in \cite{CKL}. Additionally, it has been studied in \cite{BK}. In this work, we generalize this model to incorporate a generalized complex structure, presenting it in this form. Furthermore, we provide a new proof of its integrability and introduce new examples.}
\begin{equation}\label{smg}
S=\int(l^{-1}\partial{l})^{\alpha}(g_{\alpha\beta}+kg_{\alpha\delta}J^{\delta}\hspace{0cm}_{\beta})(l^{-1}\bar{\partial}{l})^{\beta}dzd\bar{z}.
\end{equation}
Here, $g_{\alpha\beta}$ represents the metric on the Lie algebra $L$ of the Lie group $\mathbf{L}$, and $J^{\alpha}\hspace{0cm}_{\beta}$ is an endomorphism of $L$, denoted as $J: L\longrightarrow L$. The indices $\alpha, \beta, \ldots$ are Lie algebra indices. With this setup, we can repeat the calculations from the previous subsection using the following Ansatz \cite{RS1}:
\begin{equation}
\alpha_{\mu} =(\lambda_{1}J^{\alpha}\hspace{0cm}_{\beta}+\lambda_{2}\delta^{\alpha}\hspace{0cm}_{\beta})e^{\beta}\hspace{0cm}_{\mu}T_{\alpha}\hspace{0.5cm} ,\hspace{0.5cm}\mu_{\mu} =(\lambda'_{1}J^{\alpha}\hspace{0cm}_{\beta}+\lambda'_{2}\delta^{\alpha}\hspace{0cm}_{\beta})e^{\beta}\hspace{0cm}_{{\mu}}T_{\alpha},
\end{equation}
 where ${T_{\alpha}}$ are the basis of the Lie algebra $L$ with the commutation relations as
\begin{equation}\label{tt}
[T_{\alpha},T_{\beta}]=f^{\gamma}\hspace{0cm}_{\alpha\beta}T_{\gamma},
\end{equation}
$f^{\gamma}\hspace{0cm}_{\alpha\beta}$ are the structure constants of the Lie algebra that satisfy the Maurer–Cartan equation
\begin{equation}\label{mc}
f^{\gamma}\hspace{0cm}_{\alpha\beta}=e^{\gamma}\hspace{0.0cm}_{\nu}(e_{\alpha}\hspace{0.0cm}^{\mu}\partial_{\mu}e_{\beta}\hspace{0.0cm}^{\nu}-e_{\beta}\hspace{0.0cm}^{\mu}\partial_{\mu}e_{\alpha}\hspace{0.0cm}^{\nu}).
\end{equation}
Furthermore by using (\ref{gj}) the relations (\ref{jt}) and (\ref{hcsc}) can be replaced with the following algebraic form \footnote{The relations (\ref{jtg}), (\ref{hcg}) and (\ref{mcg}) show that $J^{\alpha}\hspace{0cm}_{\beta}$ is an algebraic Hermitian complex structure \cite{RA}.}

\begin{equation}\label{jtg}
J^{\alpha}\hspace{0cm}_{\beta}J^{\beta}\hspace{0cm}_{{\gamma}}=-\delta^{\alpha}\hspace{0cm}_{\gamma},
\end{equation}
\begin{equation}\label{hcg}
J^{\alpha}\hspace{0cm}_{\beta}=-g_{\beta\gamma}J^{\gamma}\hspace{0cm}_{\delta}g^{\delta\alpha},
\end{equation}
where $g_{\alpha\beta}$ is the ad-invariant metric that satisfies \cite{NW}:
\begin{equation}\label{adg}
f^{\gamma}\hspace{0cm}_{\alpha\beta}g_{\gamma\delta}+f^{\gamma}\hspace{0cm}_{\alpha\delta}g_{\gamma\beta}=0.
\end{equation}
Then using the Maurer–Cartan equation and the algebraic form of Nijenhuis condition (\ref{nj})
\begin{equation}\label{mcg}
f^{\gamma}\hspace{0cm}_{\beta\alpha}-J^{\sigma}\hspace{0cm}_{\beta}J^{\delta}\hspace{0cm}_{\alpha}f^{\gamma}\hspace{0cm}_{\sigma\delta}+J^{\gamma}\hspace{0cm}_{\sigma}J^{\delta}\hspace{0cm}_{\alpha}f^{\sigma}\hspace{0cm}_{\beta\delta}+J^{\sigma}\hspace{0cm}_{\beta}J^{\gamma}\hspace{0cm}_{\delta}f^{\delta}\hspace{0cm}_{\sigma\alpha}=0,
\end{equation}
and the Christofell and tortion relations of the model (\ref{smg})
\begin{equation}
\Gamma^{\lambda}\hspace{0cm}_{\mu\nu}=\frac{1}{2}(\partial_{\mu}e^{\alpha}\hspace{0cm}_{\nu}+\partial_{\nu}e^{\alpha}\hspace{0cm}_{\mu})e_{\alpha}\hspace{0cm}^{\lambda},
\end{equation}
\begin{equation}
 H^{\lambda}\hspace{0cm}_{\mu\nu}=\frac{k}{2}[f^{\delta}\hspace{0cm}_{\alpha\beta}J^{\gamma}\hspace{0cm}_{\delta
}+f^{\gamma}\hspace{0cm}_{\delta\alpha}J^{\delta}\hspace{0cm}_{\beta
 }+f^{\gamma}\hspace{0cm}_{\beta\delta}J^{\delta}\hspace{0cm}_{\alpha}]e^{\alpha}\hspace{0cm}_{\mu}e^{\beta}\hspace{0cm}_{\nu}e_{\gamma}\hspace{0cm}^{\lambda},
\end{equation}
after some calculation one can conclude the relation (\ref{eqn:sp}) automatically satisfies\footnote{Note that eq (\ref{eqn:sp}) show that the $J^{\alpha}\hspace{0cm}_{\beta}e^{\beta}\hspace{0cm}_{\mu}T_{\alpha}$ are killing vectors of the metric $g_{\alpha\beta}$.}, and the relation (\ref{eqn:asp}) reduces to the algebraic Nijenhuis relation (\ref{mcg}) by setting
\begin{equation}
 \lambda_{{2}}=\frac{-\lambda_{1}+k}{k},\hspace{0.5cm}\lambda'_{1}=\frac{2\lambda_{1}^2}{k-2\lambda_{1}},\hspace{0.5cm}\lambda'_{2}=\frac{2\lambda_{1}(-\lambda_{1}+k)}{k(k-2\lambda_{1})}.
\end{equation}
 In this manner, we have demonstrated that the sigma model (\ref{smg}) is integrable if and only if the endomorphism $J^{\alpha}\hspace{0cm}_{\beta}$ represents a Hermitian complex structure on $L$. Here $\lambda_{1}$ is a spectral parameter and from (\ref{lax}) we have the following Lax pairs:
\begin{align}
[\partial+\big{(}\lambda_{1}J^{\alpha}\hspace{0cm}_{\beta}+(\frac{-\lambda_{1}+k}{k}\big{)}\delta^{\alpha}\hspace{0cm}_{\beta})e^{\beta}\hspace{0cm}_{\mu}T_{\alpha}{\partial}x^{\mu}]\psi=0,
\notag\end{align}
\begin{equation}
[\bar\partial+\big{(}(\frac{k\lambda_{1}}{k-2\lambda_{1}})J^{\alpha}\hspace{0cm}_{\beta}+(\frac{-\lambda_{1}+k}{k-2\lambda_{1}})\delta^{\alpha}\hspace{0cm}_{\beta}\big{)}e^{\beta}\hspace{0cm}_{\mu}T_{\alpha}{\bar\partial}x^{\mu}]\psi=0.
\end{equation}
Note that, the matrix forms of the Lie algebra bases $T_{\alpha}$ serve as matrix representations for $\alpha_{\mu}$ and $\beta_{\mu}$.

Indeed, one can consider the above model as a special case of the generalization of the chiral model,
\begin{equation}
S(g)=\int dzd\bar{z} \Omega_{\alpha\beta}(l^{-1}\partial l)^{\alpha}(l^{-1}\bar{\partial} l)^{\beta},
\end{equation}
where $ \Omega_{\alpha\beta}$ is a constant matrix. The integrability of this model has been investigated in \cite{CI,NJ,HO} (see also \cite{NM}).
\vspace{0.5cm}

{\bf 4.1 Examples}
\vspace{0.3cm}

Now we will consider two example of the integrable sigma model (\ref{smg}) on the Lie groups $\mathbf{H_{4}}$ and $\mathbf{G_{6,23}}$.

\textbf{c)} We begin by examining an example involving the four-dimensional Heisenberg Lie group $\mathbf{H_{4}}$. The Lie algebra of this Lie group is isomorphic to $A_{4,8}$ as classified in the four-dimensional real Lie algebras \cite{PJ}. We can express the commutation relations for the $A_{4,8}$ Lie algebra as follows, based on \cite{EA}:
\begin{equation}\label{h41}
[P_{1},P_{2}]=P_{2}\hspace{0.5cm},\hspace{0.5cm}[P_{1},J]=-J\hspace{0.5cm},\hspace{0.5cm}[J,P_{2}]=T.
\end{equation}
To calculate the vierbein, we parameterize the corresponding Lie group $\mathbf{H_{4}}$ using coordinates $x^{\mu} = \{x, y, u, v\}$. In this way, its $l$ elements can be expressed as:
\begin{equation}\label{h42}
l=e^{vT_{4}}e^{uT_{3}}e^{xT_{1}}e^{yT_{2}},
\end{equation}
with the generators $T_{\alpha} = \{P_{1}, P_{2}, J, T\}$. As a result, the vierbein (\ref{vb}) and the ad-invariant metric (\ref{adg}) take the following form, as found in \cite{EA} and \cite{EP}
\begin{equation}\label{h43}
(e^{\alpha}\hspace{0.0cm}_{\mu})=\left(
\begin{array}{cccc}
  1 & 0&0&0 \\
  y &1&0&0 \\
  0&0&e^x&0\\
  0&0&y e^x&1\\
 \end{array}\right)\hspace{0.1cm},\hspace{0.1cm}(g_{\alpha\beta})=\left(
\begin{array}{cccc}
  m &0&0&-k_{0} \\
  0 &0&k_{0}&0 \\
  0&k_{0}&0&0\\
 -k_{0}&0&0&0\\
 \end{array}\right).
\end{equation}
Here, $k_{0}$ and $m$ represents an arbitrary real constant. We will now utilize a specific type of complex structure for this Lie algebra, as described in \cite{RA}
\begin{equation}
(J^{\alpha}\hspace{0.0cm}_{\beta})=\left(
\begin{array}{cccc}
  0 &0&1&0 \\
 0 &0&0&1\\
  -1&0&0&0\\
 0&-1&0&0\\
 \end{array}\right),
\end{equation}
 this structure satisfies the Hermitian complex structure (\ref{hcg}) by setting $m=0$. One can construct the action (\ref{smg}), after omitting surface terms, as
\begin{equation}\label{cse}
S=\int {dz}d\bar{z}[-k_{0}(\partial x \bar{\partial}v +\partial v \bar{\partial}x)+k_{0}e^{x}(\partial u\bar{\partial}y+\partial y\bar{\partial}u)+kk_{0}e^x(\partial u\bar{\partial}v-\partial v\bar{\partial}u)].
\end{equation}
 On the other hand, it is well-known that the $WZW$ model on a Lie group $\mathbf{L}$ is defined as follows \cite{W}:
\begin{equation}\label{wzwt}
S_{WZW}=\frac{K}{4\pi}\int_{\Sigma}dz d\bar{z}\hspace{1mm} e^{\alpha}\hspace{0cm}_{\mu}g_{\alpha \beta}e^{\beta}\hspace{0cm}_{\nu}\hspace{1mm} \partial x^{\mu}\bar{\partial}x^{\nu}+\frac{K}{24\pi}\int_{B}\hspace{2mm}d^3
\sigma\varepsilon^{i j k}e^{\alpha}\hspace{0cm}_{\mu}g_{\alpha \delta}f^{\delta}\hspace{0cm}_{\beta \gamma}e^{\beta}\hspace{0cm}_{\nu}e^{\gamma}\hspace{0cm}_{\lambda}\partial_{i}x^{\mu}\partial_{j}x^{\nu}\partial_{k}x^{\lambda},
\end{equation}
here, the world sheet $\Sigma$ serves as the boundary of a 3-dimensional bulk $B$ with coordinates $\sigma^i$. The $WZW$ model on the Heisenberg Lie group $\mathbf{H_{4}}$, as discussed in \cite{EA}, is given by the following action:
\begin{equation}\label{a48wzw}
S_{WZW_{A48}}=\int {dz}d\bar{z}[m\partial x\bar{\partial}x-k_{0}(\partial x\bar{\partial}v +\partial v\bar{\partial}x)+k_{0} e^{x}(\partial y\bar{\partial}u+\partial u\bar{\partial}y+y(\partial u\bar{\partial}x-\partial x\bar{\partial}u))],
\end{equation}
so, by assuming $K=4\pi$ and $m=0$, the action (\ref{cse}) can be consider as the perturbed $WZW$ action\footnote{Note that in equation (\ref{cse}), we have both added and subtracted the term $k_{0}ye^{x}(\partial u\bar{\partial}x-\partial x\bar{\partial}u)$ in order to construct the WZW action.}
 \begin{equation}\label{a48wzwd}
 S=S_{WZW_{A_{4,8}}}+\int {dz}d\bar{z}[-k_{0}ye^{x}(\partial u\bar{\partial}x-\partial x\bar{\partial}u)+kk_{0}e^x(\partial u\bar{\partial}v-\partial v\bar{\partial}u)].
\end{equation}
 Therefore, the integrable sigma model with a complex structure on the Heisenberg Lie group $\mathbf{H_{4}}$ is equivalent to the integrable perturbed $WZW$ sigma model. It's worth noting that our model (\ref{cse}) is the same as  Yang-Baxter deformation of the $WZW$ model of Heisenberg Lie group $\mathbf{H}_{4}$ (case $VII$ with $\rho=\eta=0$) of table 1 of ref \cite{EP}.

\vspace{0.5cm}
\textbf{d)} As another example, we construct (\ref{smg}) model on six-dimensional Lie group $\mathbf{G_{6,23}}$. The $g_{6,23}$ Lie algebra have the following commutations \cite{MO,RS}:
\begin{equation}
[T_{2},T_{3}]=T_{1}\hspace{0.5cm},\hspace{0.5cm}[T_{2},T_{6}]=T_{3}\hspace{0.5cm},\hspace{0.5cm}[T_{3},T_{6}]=T_{4}.
\end{equation}
Now using representation $l=e^{x_{1}T_{1}}e^{x_{2}T_{2}}e^{x_{3}T_{3}}e^{x_{4}T_{4}}e^{x_{5}T_{5}}e^{x_{6}T_{6}}$ for a Lie group element with coordinates $x^{\mu}=\{x_{1},x_{2},x_{3},x_{4},x_{5},x_{6}\}$ and the generators $\{T_{1}, T_{2}, T_{3}, T_{4}, T_{5}, T_{6}\}$, the algebraic  metric and veilbeins related to this Lie algebra are given as \cite{RS}:
\begin{equation}\label{ec6}
(e^{\alpha}\hspace{0.0cm}_{\mu})=\left(
\begin{array}{cccccc}
  1 & x_{3}&0&0&0&0 \\
  0 &1&0&0&0&0 \\
  0&x_{6}&1&0&0&0\\
  0&\frac{1}{2}x_{6}^{2}&x_{6}&1&0&0\\
  0 &0&0&0&1&0 \\
  0 &0&0&0&0&1 \\
 \end{array}\right)\hspace{0.1cm},\hspace{0.1cm}(g_{\alpha\beta})=\left(
\begin{array}{cccccc}
  0 &0&0&0&0&m_{2} \\
  0 &m_{1}&0&m_{2}&m_{3}&m_{4} \\
  0&0&-m_{2}&0&0&0\\
  0&m_{2}&0&0&0&0\\
  0 &m_{3}&0&0&m_{5}&m_{6} \\
  m_{2} &m_{4}&0&0&m_{6}&m_{7} \\
\end{array}\right),
\end{equation}
where $m_{1},..., m_{7}$ are arbitrary real parameters. By utilizing equations (\ref{jtg} - \ref{mcg}) and the metric (\ref{ec6}), and after some calculations, we can determine the algebraic integrable Hermitian complex structure $J$ as follows:
\begin{equation}
(J^{\alpha}\hspace{0.0cm}_{\beta})=\left(
\begin{array}{cccccc}
  -a & 0&0&\frac{a^2+1}{b}&0&0 \\
  0 &-a&0&0&0&-\frac{a^2+1}{b} \\
  0&0&0&0&-\frac{1}{c}&0\\
  -b&0&0&a&0&0\\
  0 &0&c&0&0&0 \\
  0 &b&0&0&0&a \\
 \end{array}\right),
\end{equation}
where $a\in\Re$, $b, c\in\Re-\{0\}$ are arbitrary real parameters and $m_{3}=0,m_{4}=\frac{am_{1}}{b},m_{5}=-\frac{m_{2}}{c^2},m_{6}=0,m_{7}=\frac{(a^2+1)m_{1}}{b^2}$. Thus, our two-dimensional integrable sigma model (\ref{smg}) can be expressed as follows (after omitting surface terms):
\begin{equation*}
  S =\int dzd\bar{z}[m_{1}(\partial x_{2}\bar{\partial}x_{2}+\frac{(a^2+1)}{b^2}\partial x_{6}\bar{\partial}x_{6})+m_{2}(-\partial x_{3}\bar{\partial}x_{3}- \frac{1}{c^2}\partial x_{5}\bar{\partial}x_{5}+\partial x_{6}\bar{\partial}x_{1}+\partial x_{1}\bar{\partial}x_{6}
+\partial x_{4}\bar{\partial}x_{2}+\partial x_{2}\bar{\partial}x_{4})
 \notag\end{equation*}
\begin{equation}\label{g623cm}
+\frac{b m_{2}x_{3}+am_{1}}{b}(\partial x_{6}\bar{\partial}x_{2}+\partial x_{2}\bar{\partial}x_{6})
-\frac{m_{2}}{c}kx_{6}(\partial x_{5}\bar{\partial}x_{2}-\partial x_{2}\bar{\partial}x_{5})],
\end{equation}
one can compare this model with $WZW$ model (\ref{wzwt}) on $\mathbf{g_{6,23}}$ which is given by :
\begin{equation*}
  S_{WZW_{g_{6,23}}} =\frac{K}{4\pi}\int dzd \bar{z}[m_{1}(\partial x_{2}\bar{\partial}x_{2}+\frac{(a^2+1)}{b^2}\partial x_{6}\bar{\partial}x_{6})+m_{2}(-\partial x_{3}\bar{\partial}x_{3}- \frac{1}{c^2}\partial x_{5}\bar{\partial}x_{5}+\partial x_{6}\bar{\partial}x_{1}+\partial x_{1}\bar{\partial}x_{6}
 \notag \end{equation*}
\begin{equation}
+\partial x_{4}\bar{\partial}x_{2}+\partial x_{2}\bar{\partial}x_{4})+\frac{b m_{2}x_{3}+am_{1}}{b}(\partial x_{6}\bar{\partial}x_{2}+\partial x_{2}\bar{\partial}x_{6})-m_{2}x_{3}(\partial x_{2}\bar{\partial}x_{6}-\partial x_{6}\bar{\partial}x_{2})].
\end{equation}
Indeed, the model (\ref{g623cm}) can be considered as a perturbation of the $WZW$ model with the following perturbed term, achieved by setting $K = 4\pi$ :

\begin{equation}
  S =S_{WZW_{g_{6,23}}} + \int dzd\bar{z}[m_{2}x_{6}(\partial x_{3}\bar{\partial}x_{2}-\partial x_{2}\bar{\partial}x_{3})-\frac{m_{2}}{c}kx_{6}(\partial x_{5}\bar{\partial}x_{2}-\partial x_{2}\bar{\partial}x_{5})].
\end{equation}
 In addition, one can investigate the conformality of this model up to one loop, where the one-loop $\beta$ function equations \cite{CC} are as follows:
\begin{equation*}
 B^{g}\hspace{0.0cm}_{\mu\nu}=-\alpha^{'}[R_{\mu\nu}-(H^{2})_{\mu\nu}+\nabla_{\mu}\nabla_{\nu}\phi]=0,
 \notag \end{equation*}\begin{equation*}
B^{B}\hspace{0.0cm}_{\mu\nu}=-\alpha^{'}[-\nabla^{\lambda}H_{\lambda\mu\nu}+H_{\mu\nu}\hspace{0.0cm}^{\lambda}\nabla_{\lambda}\phi]=0,
 \notag \end{equation*}
 \begin{equation}\label{bf}
 B^{\phi}\hspace{0.0cm}_{\mu\nu}=-\alpha^{'}[R-\frac{1}{3}H^2-\frac{1}{2}\nabla^{2}\phi+\frac{1}{2}(\nabla\phi)^2]=0,
\end{equation}
where $H^{2}\hspace{0.0cm}_{\mu\nu}=H_{\mu\rho\lambda}H^{\rho\lambda}\hspace{0.0cm}_{\nu}$ and $H^2=H_{\mu\rho\lambda}H^{\mu\rho\lambda}$. The $Ricci$ components $(R_{\mu\nu})$ of this sigma model are zero, and the torsion components are given as follows:
\begin{equation}
H^{1}\hspace{0.0cm}_{25}=\frac{k}{2c}\hspace{0.2cm},\hspace{0.2cm}H^{4}\hspace{0.0cm}_{56}=\frac{k}{2c}\hspace{0.2cm},\hspace{0.2cm}H^{5}\hspace{0.0cm}_{26}=\frac{kc}{2}.
\end{equation}
So, all of the one-loop $\beta$ function equations are satisfied with $\phi=\text{const}$. In the following section, we will generalize our model (\ref{smg}) for the complex structure to a model with a generalized complex structure.

\section{\bf  Integrable sigma model with generalized complex structure}\vspace{0.5cm}\label{sec5}
Let's first have a short review of concepts and notations related to generalized complex structure \cite{HI},\cite{GU}.
\subsection{\bf  Review of generalized complex structure}\vspace{0.5cm}
A generalized complex structure on a manifold $M$ (with even dimension) is an endomorphism $\mathcal{J}: TM\oplus T^*M \rightarrow TM\oplus T^*M$ such that $\mathcal{J}$ is invariant with respect to the inner product $\langle\ ,\ \rangle$ on $TM\oplus T^*M$.
\begin{equation}
\forall X,Y \in {TM}\hspace{0.1cm}, \hspace{0.1cm} \xi,\eta \in {{T}^*{M}}:\hspace{1cm}\langle {\mathcal{J}}(X+\xi) , \mathcal{J}(Y+\eta)\rangle=\langle
X+\xi,Y+\eta\rangle ,
\end{equation}
and $\mathcal{J}^{2}=-1$.
Using the Courant bracket on a smooth section of
${TM}\oplus{T^*M}$ \cite{CU}
\begin{equation}\label{cua}
[X+\xi , Y+\eta]_{C}=[X,Y]+L_{X}{\eta}-L_{Y}{\xi}-\dfrac{1}{2}d_M(i_{X}\eta-i_{Y}\xi),
\end{equation}
generalized complex structure $\mathcal{J}$ is an integrable structure if the generalized Nijenhuis tensor is zero \cite{HI,GU}
\begin{equation}\label{nij}
N_\mathcal{J}(X+\xi,Y+\eta)=[X+\xi,Y+\eta]_{C}+\mathcal{J}[X+\xi,\mathcal{J}(Y+\eta)]_{C}+\mathcal{J}[\mathcal{J}(X+\xi),(Y+\eta)]_{C}-[\mathcal{J}(X+\xi),\mathcal{J}(Y+\eta)]_{C}=0.
\end{equation}
 We can consider the almost generalized complex structure in the following block form, as discussed in \cite{GU}:
\begin{equation}
\mathcal{J}=\left(
\begin{array}{cccc}
  J & P \\
  Q &-J^* \\
 \end{array}\right),
\end{equation}
 where $J=J^{\mu}\hspace{0cm}_{\nu}\partial_{\mu}\otimes dx^{\nu}$,
 $P=P^{\mu\nu}\partial_{\mu}\wedge \partial_{\nu}$
 and $Q=Q_{\mu\nu}dx^{\mu}\wedge dx^{\nu}$. By applying the above block form to $\mathcal{J}^2=-1$, we obtain the following relations for the tensors $J$, $P$, and $Q$:
\begin{equation}\label{gcs1}
P^{\nu\\k}+P^{k\nu}=0\hspace{0.5cm} ,\hspace{0.5cm}Q_{\nu\\k}+Q_{k\nu}=0,
\end{equation}
\begin{equation}
J^{\nu}\hspace{0cm}_{\mu}J^{\mu}\hspace{0cm}_{k}+P^{\nu\mu}Q_{{\mu}{k}}+\delta^{{\nu}}\hspace{0cm}_{k}=0,
\end{equation}
\begin{equation}
J^{\nu}\hspace{0cm}_{\mu}P^{\mu\\k}+J^{k}\hspace{0cm}_{\mu}P^{\mu\nu}=0 ,
\end{equation}
\begin{equation}\label{gcs4}
Q_{\nu\mu}J^{\mu}\hspace{0cm}_{k}+Q_{k\mu}J^{\mu}\hspace{0cm}_{\nu}=0.
\end{equation}
Furthermore, by using the Courant bracket definition (\ref{cua}), the integrability condition of the generalized complex structure (\ref{nij}) can be expressed as the following tensor relations, as shown in \cite{RZ}:
\vspace{0.02cm}
\begin{equation}\label{njg1}
\mathbf{A}^{\nu\\k\mu}=P^{\nu\lambda}\partial_{\lambda}P^{k\mu}+P^{k\lambda}\partial_{\lambda}P^{\mu\nu}+P^{\mu\lambda}\partial_{\lambda}P^{\nu\\k}=0,
\end{equation}
\begin{equation}\label{njg2}
\mathbf{B}^{k\mu}\hspace{0cm}_{\nu}=J^{\lambda}\hspace{0cm}_{\nu}\partial_{\lambda}P^{k\mu}+P^{k\lambda}(\partial_{\nu}J^{\mu}\hspace{0cm}_{\lambda}-\partial_{\lambda}J^{\mu}\hspace{0cm}_{\nu})-P^{\mu\lambda}(\partial_{\nu}J^{k}\hspace{0cm}_{\lambda}-\partial_{\lambda}J^{k}\hspace{0cm}_{\nu})-\partial_{\nu}(J^{k}\hspace{0cm}_{\lambda}P^{\lambda\mu})=0,
\end{equation}
\begin{equation}\label{njg3}
\mathbf{C}^{\mu}\hspace{0cm}_{\nu\\k}=J^{\lambda}\hspace{0cm}_{\nu}\partial_{\lambda}J^{\mu}\hspace{0cm}_{k}-J^{\lambda}\hspace{0cm}_{k}\partial_{\lambda}J^{\mu}\hspace{0cm}_{\nu}-J^{\mu}\hspace{0cm}_{\lambda}\partial_{\nu}J^{\lambda}\hspace{0cm}_{k}+J^{\mu}\hspace{0cm}_{\lambda}\partial_{k}J^{\lambda}\hspace{0cm}_{\nu}+P^{\mu\lambda}(\partial_{\lambda}Q_{\nu\\k}+\partial_{\nu}Q_{k\lambda}+\partial_{k}Q_{\lambda\nu})=0,
\end{equation}
\begin{equation*}
  \mathbf{D}_{\nu k
  \mu}=J^{\lambda}\hspace{0cm}_{\nu}(\partial_{\lambda}Q_{k\mu}+\partial_{k}Q_{\mu\lambda}+\partial_{\mu}Q_{\lambda
  k})+J^{\lambda}\hspace{0cm}_{k}(\partial_{\lambda}Q_{\mu\nu}+\partial_{\mu}Q_{\nu\lambda}+\partial_{\nu}Q_{\lambda\mu})+J^{\lambda}\hspace{0cm}_{\mu}(\partial_{\lambda}Q_{\nu
  k}+\partial_{\nu}Q_{k\lambda}+\partial_{k}Q_{\lambda\nu})
\notag\end{equation*}
\begin{equation}\label{njg4}
-\partial_{\nu}(Q_{k\lambda}J^{\lambda}\hspace{0cm}_{\mu})-\partial_{k}(Q_{\mu\lambda}J^{\lambda}\hspace{0cm}_{\nu})-\partial_{\mu}(Q_{\nu\lambda}J^{\lambda}\hspace{0cm}_{k})=0.
\end{equation}
These relations are necessary conditions for the integrability of generalized complex structure.
\subsection{\bf Construction of the integrable sigma model}

\subsubsection{\bf Model on manifold}
Now, by using the components of the generalized complex structure on $TM$ and $T^{*}M$, we propose the following sigma model action on the manifold $M$ with coordinates $x^{\mu}$ and metric $g_{\mu\nu}$:
\begin{equation}\label{gcsm}
S=\int\hspace{2mm}dzd\bar{z}(g_{\mu\nu}+k
g_{\mu\lambda}J^{\lambda}\hspace{0mm}_{\nu}+k^{'}Q_{\mu\nu}+k^{''}g_{\mu\lambda}P^{\lambda\gamma}g_{\gamma\nu}
)\hspace{0mm}\partial\hspace{0mm}x^{\mu}\bar{\partial}x^{\nu},
\end{equation}
where $k$ , $k^{'}$ and $k^{''}$ are non-zero constants. Note that by comparing (\ref{gcsm}) with (\ref{eqn:sm}), the components $G_{\mu\nu}$ and $B_{\mu\nu}$ (corresponding to the symmetric and antisymmetric parts) of this model have the following form:
\begin{equation}
G_{\mu\nu}=g_{\mu\nu}+\dfrac{k}{2}(g_{\mu\lambda}J^{\lambda}\hspace{0cm}_{\nu}+g_{\nu\lambda}J^{\lambda}\hspace{0cm}_{\mu}),
\end{equation}
\begin{equation}
B_{\mu\nu}=\dfrac{k}{2}(g_{\mu\lambda}J^{\lambda}\hspace{0cm}_{\nu}-g_{\nu\lambda}J^{\lambda}\hspace{0cm}_{\mu})+k^{'}Q_{\mu\nu}+k^{''}g_{\mu\lambda}P^{\lambda\gamma}g_{\gamma\nu}.
\end{equation}
The metric $G_{\mu\nu}$ must be invertible, which means $G^{\mu\lambda}G_{\lambda\nu}=\delta^{\mu}\hspace{0cm}_{\nu}$. So by assuming $J^{\mu}\hspace{0cm}_{\nu}$ satisfies the Hermitian complex structure
(\ref{hcsc}) then $G_{\mu\nu}$ and $B_{\mu\nu}$ have the following form:
 \begin{equation}\label{gbgc}
 G_{\mu\nu}=g_{\mu\nu}\hspace{1.5cm},\hspace{1.5cm}B_{\mu\nu}=k
 g_{\mu\lambda}J^{\lambda}\hspace{0cm}_{\nu}+k^{'}Q_{\mu\nu}+k^{''}g_{\mu\lambda}P^{\lambda\gamma}g_{\gamma\nu}.
 \end{equation}
Now, by using relations (\ref{gcs1}-\ref{njg4}), one can derive conditions on the tensors $J$, $P$, and $Q$. To ensure the integrability of the model (\ref{gcsm}), one must further impose condition (\ref{me}) (or conditions (\ref{eqn:sp}) and (\ref{eqn:asp})) on (\ref{gbgc}). When $M$ is a Lie group, these conditions are relatively simple compared to the general manifold $M$. Therefore, we consider the model on a Lie group.
 \subsubsection{\bf Model on Lie group}

 In the case where $M$ is a Lie group $\mathbf{L}$, using the veilbein formalism
  \begin{equation}
{{\forall}{l}\in\mathbf{L}}\hspace{0.5cm},\hspace{0.5cm}(l^{-1}\partial{l})^{\alpha}=e^{\alpha}\hspace{0cm}_{\mu}{\partial}x^{\mu},
\end{equation}
the algebraic structure (\ref{gj}) and
\begin{equation}
P^{\mu\nu}=e_{\alpha}\hspace{0cm}^{\mu}P^{\alpha\beta}e_{\beta}\hspace{0cm}^{\nu}\hspace{1cm},\hspace{1cm} Q_{\mu\nu}=e^{\alpha}\hspace{0cm}_{\mu}Q_{\alpha\beta}e^{\beta}\hspace{0cm}_{\nu},
\end{equation}
the action (\ref{gcsm}) can be rewritten as
\begin{equation}\label{gcsm1}
S=\int dzd\bar{z}(l^{-1}\partial l)^{\alpha}(g_{\alpha\beta}+k
g_{\alpha\delta}J^{\delta}\hspace{0cm}_{\beta}+k^{'}Q_{\alpha\beta}+k^{''}g_{\alpha\delta}P^{\delta\sigma}g_{\sigma\beta})(l^{-1}\bar{\partial}
l)^{\beta}.
\end{equation}
Now, by applying the formalism from reference \cite{NM}, as discussed in section \ref{sec2}, one can examine the integrability conditions of this sigma model. For this model, the Christoffel and torsion are given by:
\begin{equation}\label{gama}
 \Gamma^{\lambda}\hspace{0cm}_{\mu\nu}=\frac{1}{2}(\partial_{\mu}e^{\alpha}\hspace{0cm}_{\nu}+\partial_{\nu}e^{\alpha}\hspace{0cm}_{\mu})e_{\alpha}\hspace{0cm}^{\lambda},
\end{equation}
\begin{equation*}
 H^{\lambda}\hspace{0cm}_{\mu\nu}=\frac{1}{2}[k(f^{\alpha}\hspace{0cm}_{\delta\gamma}J^{\delta}\hspace{0cm}_{\beta
}-f^{\alpha}\hspace{0cm}_{\delta\beta}J^{\delta}\hspace{0cm}_{\gamma
 }-f^{\delta}\hspace{0cm}_{\beta\gamma}J^{\alpha}\hspace{0cm}_{\delta})+k'(f^{\delta}\hspace{0cm}_{\gamma\sigma}Q_{\delta\beta}g^{\alpha\sigma}+f^{\delta}\hspace{0cm}_{\sigma\beta}Q_{\delta\gamma}g^{\alpha\sigma}-f^{\delta}\hspace{0cm}_{\gamma\beta}Q_{\delta\sigma}g^{\alpha\sigma})
 \notag\end{equation*}
 \begin{equation}\label{Hgcm}
  +k''(f^{\alpha}\hspace{0cm}_{\delta\gamma}P^{\delta\sigma}g_{\sigma\beta}-f^{\alpha}\hspace{0cm}_{\delta\beta}P^{\delta\sigma}g_{\sigma\gamma}-f^{\delta}\hspace{0cm}_{\gamma\beta}P^{\sigma\alpha}g_{\delta\sigma})]e^{\gamma}\hspace{0cm}_{\mu}e^{\beta}\hspace{0cm}_{\nu}e_{\alpha}\hspace{0cm}^{\lambda}.
  \end{equation}
 We assume $\alpha_{\mu}$ and $\mu_{\mu}$ have the following forms:
 \begin{equation*}
{\alpha}_{\mu}=(\lambda_{1}J^{\alpha}\hspace{0cm}_{\gamma}+\lambda_{2}P^{\alpha\delta}g_{\delta\gamma}+\lambda_{3}g^{\alpha\delta}Q_{\delta\gamma}+\lambda_{4}\delta^{\alpha}\hspace{0cm}_{\gamma})e^{\gamma}\hspace{0cm}_{\mu}T_{\alpha},
\notag\end{equation*}
\begin{equation}\label{am}
{\mu}_{\mu}=(\lambda'_{1}J^{\alpha}\hspace{0cm}_{\gamma}+
\lambda'_{2}P^{\alpha\delta}g_{\delta\gamma}+\lambda'_{3}g^{\alpha\delta}Q_{\delta\gamma}+\lambda'_{4}\delta^{\alpha}\hspace{0cm}_{\gamma})e^{\gamma}\hspace{0cm}_{\mu}T_{\alpha},
 \end{equation}
where $\{\lambda_{1}, \lambda_{2}, \lambda_{3}, \lambda_{4}, \lambda'_{1}, \lambda'_{2}, \lambda'_{3}, \lambda'_{4}\}$ are arbitrary parameters, and $T_{\alpha}$ represents the bases of the Lie algebra $L$, satisfying the commutation relations (\ref{tt}). Furthermore one can rewrite the conditions for the generalized complex structure (\ref{gcs1}-\ref{gcs4}) and its integrability (\ref{njg1}-\ref{njg4}) as the following algebraic relations \cite{FRD}:
\begin{equation}\label{pp}
P^{\alpha\beta}+P^{\beta\alpha}=0\hspace{0.5cm} ,\hspace{0.5cm}Q_{\alpha\beta}+Q_{\beta\alpha}=0,
\end{equation}
\begin{equation}\label{jj}
J^{\alpha}\hspace{0cm}_{\delta}J^{\delta}\hspace{0cm}_{\beta}+P^{\alpha\delta}Q_{\delta\beta}+\delta^{\alpha}\hspace{0cm}_{\beta}=0
,
\end{equation}
\begin{equation}\label{jp}
J^{\alpha}\hspace{0cm}_{\delta}P^{\delta\beta}+J^{\beta}\hspace{0cm}_{\delta}P^{\delta\alpha}=0,
\end{equation}
\begin{equation}\label{qjeq}
Q_{\alpha\delta}J^{\delta}\hspace{0cm}_{\beta}+Q_{\beta\delta}J^{\delta}\hspace{0cm}_{\alpha}=0,
\end{equation}
\begin{equation}\label{A}
\mathbf{A^{\alpha\beta\gamma}}=f^{\alpha}\hspace{0cm}_{\delta\sigma}P^{\beta\sigma}P^{\gamma\delta}+f^{\gamma}\hspace{0cm}_{\delta\sigma}P^{\beta\delta}P^{\alpha\sigma}+f^{\beta}\hspace{0cm}_{\delta\sigma}P^{\alpha\delta}P^{\gamma\sigma}=0
, \end{equation}
\begin{equation}\label{B}
\mathbf{B^{\beta\gamma}\hspace{0cm}_{\alpha}}=f^{\delta}\hspace{0cm}_{\sigma\alpha}P^{\beta\sigma}J^{\gamma}\hspace{0cm}_{\delta}+f^{\delta}\hspace{0cm}_{\alpha\sigma}P^{\gamma\sigma}J^{\beta}\hspace{0cm}_{\delta}+f^{\gamma}\hspace{0cm}_{\sigma\delta}P^{\beta\delta}J^{\sigma}\hspace{0cm}_{\alpha}+f^{\beta}\hspace{0cm}_{\sigma\delta}P^{\gamma\sigma}J^{\delta}\hspace{0cm}_{\alpha}=0
,
\end{equation}
\begin{equation}\label{C}
\mathbf{C^{\alpha}\hspace{0cm}_{\beta\gamma}}=f^{\alpha}\hspace{0cm}_{\beta\gamma}-f^{\alpha}\hspace{0cm}_{\delta\sigma}J^{\delta}\hspace{0cm}_{\beta}J^{\sigma}\hspace{0cm}_{\gamma}-f^{\delta}\hspace{0cm}_{\gamma\sigma}J^{\alpha}\hspace{0cm}_{\delta}J^{\sigma}\hspace{0cm}_{\beta}+f^{\delta}\hspace{0cm}_{\beta\sigma}J^{\alpha}\hspace{0cm}_{\delta}J^{\sigma}\hspace{0cm}_{\gamma}+f^{\delta}\hspace{0cm}_{\sigma\beta}P^{\alpha\sigma}Q_{\delta\gamma}+f^{\delta}\hspace{0cm}_{\gamma\sigma}P^{\alpha\sigma}Q_{\delta\beta}=0
,
\end{equation}
\begin{equation}\label{D}
\mathbf{D_{\alpha\beta\gamma}}=f^{\delta}\hspace{0cm}_{\alpha\sigma}J^{\sigma}\hspace{0cm}_{\beta}Q_{\delta\gamma}+f^{\delta}\hspace{0cm}_{\gamma\sigma}J^{\sigma}\hspace{0cm}_{\beta}Q_{\alpha\delta}+f^{\delta}\hspace{0cm}_{\gamma\sigma}J^{\sigma}\hspace{0cm}_{\alpha}Q_{\delta\beta}+f^{\delta}\hspace{0cm}_{\beta\sigma}J^{\sigma}\hspace{0cm}_{\alpha}Q_{\gamma\delta}+f^{\delta}\hspace{0cm}_{\beta\sigma}J^{\sigma}\hspace{0cm}_{\gamma}Q_{\delta\alpha}+f^{\delta}\hspace{0cm}_{\alpha\sigma}J^{\sigma}\hspace{0cm}_{\gamma}Q_{\beta\delta}=0.
\end{equation}
As an example, we have explained the derivation of equation (\ref{C}) in the Appendix. Here, $\mathbf{C^{\alpha}\hspace{0cm}_{\beta\gamma}}$ represents the generalized Nijenhuis equation. Let us examine these conditions for some examples.

\vspace{0.5cm}

\textbf{Example e)}
 As our first example, we consider the generalized complex structure on the four-dimensional real Lie group $\mathbf{A_{4,8}}$. Here we present two examples of this Lie group. By using (\ref{pp}-\ref{D}) and Hermitian complex structure (\ref{hcg}), one can derive the first algebraic forms as follows\cite{FRD}:
 \begin{equation}
i)\hspace{2cm}(J^{\alpha}\hspace{0.0cm}_{\beta})=\left(
\begin{array}{cccc}
  0 & 0&1&0 \\
  0 &0&0&1 \\
  -1&0&0&0\\
  0&-1&0&0\\
 \end{array}\right)\hspace{0.1cm},\hspace{0.1cm}(Q_{\alpha\beta})=\left(
\begin{array}{cccc}
  0 &-1&0&1 \\
  1 &0&-1&0 \\
  0&1&0&1\\
 -1&0&-1&0\\
 \end{array}\right),
 \hspace{0.1cm}
 (P^{\alpha\beta})=0,
 \end{equation}

 then, by utilizing (\ref{h43}) with $m=0 $, one can construct the integrable sigma model (\ref{gcsm1}), after omitting surface terms, as follows:
 \begin{equation*}
S_{i}=\int dzd\bar{z}[-k_{0}(\partial v\bar{\partial}x+\partial x\bar{\partial}v)+k_{0}e^{x}(\partial u\bar{\partial}y+\partial y\bar{\partial}u)-k'ye^{x}(\partial u\bar{\partial}x-\partial x\bar{\partial}u)
\end{equation*}
\begin{equation}\label{sa481}
+(k'+kk_{0})e^{x}(\partial u\bar{\partial}v-\partial v\bar{\partial}u)].
\end{equation}
The second generalized complex structure on $\mathbf{A_{4,8}}$ Lie group is given by the following algebra forms
\begin{equation}
ii) \hspace{2cm}(J^{\alpha}\hspace{0.0cm}_{\beta})=\left(
\begin{array}{cccc}
  0 & 1&0&0 \\
  -1&0&0&0 \\
  0&0&0&1\\
  0&0&-1&0\\
 \end{array}\right)\hspace{0.1cm},\hspace{0.1cm}(Q_{\alpha\beta})=\left(
\begin{array}{cccc}
  0 &0&-1&1 \\
  0 &0&1&1 \\
  1&-1&0&0\\
 -1&-1&0&0\\
 \end{array}\right),
 \hspace{0.1cm}
 (P^{\alpha\beta})=0,
 \end{equation}
 the action (\ref{gcsm1}) of this structure can be constructed as
  \begin{equation*}
S_{ii}=\int dzd\bar{z}[-k_{0}(\partial v\bar{\partial}x+\partial x\bar{\partial}v)+k_{0}e^{x}(\partial u\bar{\partial}y+\partial y\bar{\partial}u)-e^{x}(k'y+\frac{1}{2}(k'+kk_{0})y^2)(\partial u\bar{\partial}x-\partial x\bar{\partial}u)
\end{equation*}
\begin{equation}\label{sa482}
+(k'+kk_{0})y(\partial x\bar{\partial}v-\partial v\bar{\partial}x)].
\end{equation}

To investigate the integrability of the above sigma models with generalized complex structure (99) and metric $g_{\alpha\beta}$ (\ref{h43}), we consider the Christoffel  components of models (\ref{sa481}) and (\ref{sa482})
\begin{align}
\Gamma^{2}\hspace{0cm}_{12}=\frac{1}{2},\hspace{0.5cm} \Gamma^{3}\hspace{0cm}_{13}=\frac{1}{2},\hspace{0.5cm}\Gamma^{4}\hspace{0cm}_{23}=\frac{e^x}{2},
\notag\end{align}
such that torsion components of the model (\ref{sa481}) are given as
\begin{align}
H^{1}\hspace{0cm}_{31}=\frac{e^x(k'+kk_{0})}{2k_{0}},\hspace{0.5cm}  H^{2}\hspace{0cm}_{21}=\frac{ k' }{2k_{0}}
,\hspace{0.5cm} H^{2}\hspace{0cm}_{41}=\frac{k'+kk_{0}}{2k_{0}},\notag\end{align}
\begin{equation}\label{to1}
H^{3}\hspace{0cm}_{13}=\frac{k'}{2k_{0}},\hspace{0.5cm} H^{4}\hspace{0cm}_{23}=\frac{e^x k'}{2k_{0}},\hspace{0.5cm} H^{4}\hspace{0cm}_{43}=\frac{e^x(k'+kk_{0})}{2k_{0}},
 \end{equation}
 and for torsion components of the model (\ref{sa482}) we have
 \begin{align}
H^{1}\hspace{0cm}_{12}=\frac{(k'+kk_{0})}{2k_{0}},\hspace{0.5cm}  H^{2}\hspace{0cm}_{21}=\frac{ k'+y(k'+kk_{0})}{2k_{0}}
,\hspace{0.5cm}H^{3}\hspace{0cm}_{13}=\frac{k'+y(k'+kk_{0})}{2k_{0}},\notag\end{align}
\begin{equation}\label{to2}
H^{3}\hspace{0cm}_{14}=\frac{(k'+kk_{0})e^{-x}}{2k_{0}},\hspace{0.5cm}H^{4}\hspace{0cm}_{23}=\frac{(k'+y(k'+kk_{0}))e^{x}}{2k_{0}},\hspace{0.5cm}H^{4}\hspace{0cm}_{24}=\frac{(k'+kk_{0})}{2k_{0}}.
 \end{equation}
Then by using the matrices $\alpha_{\mu}$ and $\mu_{\mu}$ from the relations (\ref{am}), the integrability condition (\ref{me}) of models (\ref{sa481}) and (\ref{sa482}) is met by setting
with the following conditions:
\begin{equation}\label{a48gcc}
\lambda_{1}= \frac{(kk_{0}+k'-2\lambda_{3})\lambda'_{3}-2\lambda_{3}^{2}}{2k_{0}(\lambda_{3}+\lambda'_{3})},\hspace{0.2cm} \lambda_{4}=\frac{(k_{0}+k'-2\lambda_{3})\lambda'_{3}-2\lambda_{3}(\lambda_{3}-k_{0})}{2k_{0}(\lambda_{3}+\lambda'_{3})},\hspace{0.2cm} \lambda'_{1}=\frac{\lambda_{1}}{\lambda_{3}}\lambda'_{3},\hspace{0.2cm} \lambda'_{4}=\frac{\lambda_{4}}{\lambda_{3}}\lambda'_{3},
\end{equation}
where $\lambda_{3},\lambda'_{3}$ are arbitrary constants, and one of them can be considered as spectral parameter.
Furthermore, by comparing this model with the $WZW$ model on $H_{4}$ \cite{EA} and recalling $K=4\pi$, the actions (\ref{sa481}) and  (\ref{sa482}) correspond to the $WZW$ action (\ref{a48wzw}) with $m=0$, $k'=-kk_{0}$ and $k=1$ .i.e in these case the perturbed term is the same as $WZ$ term. For the $k'\neq-kk_{0} $ case, the models (\ref{sa481}) and  (\ref{sa482}) are integrable perturbed $WZW$ model as follows:
\begin{equation}
S_{i}=S_{WZW_{A_{4,8}}} +\int dzd\bar{z}[(k'+k_{0})ye^{x}(\partial x\bar{\partial}u-\partial u\bar{\partial}x)+(k'+kk_{0})e^{x}(\partial u\bar{\partial}v-\partial v\bar{\partial}u)].
\end{equation}
\begin{equation}
S_{ii}=S_{WZW_{A_{4,8}}} +\int dzd\bar{z}[((k'+k_{0})y+\frac{1}{2}(k'+kk_{0})y^2)e^{x}(\partial x\bar{\partial}u-\partial u\bar{\partial}x)+(k'+kk_{0})e^{x}(\partial u\bar{\partial}v-\partial v\bar{\partial}u)].
\end{equation}

Note that for $k^{'}=-k_0$ and $k=1$, the perturbed models $S_i$ and $S_{ii}$ are the same as $WZW$ model.

\textbf{Example f})
 As another example, we construct the generalized complex structure on the four-dimensional real Lie group $\mathbf{A_{4,10}}$ \cite{PJ}, which is known as the Nappi-Witten Lie group \cite{NW}. This group is characterized by the following Lie algebra commutators:
\begin{equation}
 [J,P_{1}]=P_{2}\hspace{0.5cm} ,\hspace{0.5cm}[J,P_{2}]=-P_{1}\hspace{0.5cm} ,\hspace{0.5cm}
 [P_{1},P_{2}]=T.
 \end{equation}
This Lie algebra is a generalization of the 2D Poincare algebra, which corresponds to the case when $T=0$. We adopt the following parametrization for the elements of the corresponding Lie group $\mathbf{A_{4,10}}$ \cite{NW}:
\begin{equation}
l=e^{\Sigma a_{i}P_{i}}e^{uJ+vT}\hspace{0cm},
 \end{equation}
with coordinates $x^{\mu}=\{ a_{1} ,a_{2}, u,v\}$ and the generators $T_{\alpha}=\{ P_{1}, P_{2}, J,T\}$. Then, the veirbein and ad-invariant metric (\ref{adg}) for
this Lie algebra are as follows:
\begin{equation}\label{ga410}
(e^{\alpha}\hspace{0.0cm}_{\mu})=\left(
\begin{array}{cccc}
  cos(u)&sin(u)&0&0 \\
 -sin(u)&cos(u)&0&0\\
  0&0&1&0\\
  \frac{1}{2}a_{2}&-\frac{1}{2}a_{1}&0&1 \\
 \end{array}\right)\hspace{0.1cm},\hspace{0.1cm}(g_{\alpha\beta})=\left(
\begin{array}{cccc}
  1 &0&0&0 \\
  0 &1&0&0 \\
  0&0&k_{0}&1\\
 0&0&1&0\\
 \end{array}\right),
 \end{equation}
where $k_{0}$ is a real constant. The corresponding components of the generalized complex structure, which satisfies the relations (\ref{pp}-\ref{D}) and metric compatible complex structure (\ref{hcg}) are as follows
\begin{equation}
(J^{\alpha}\hspace{0.0cm}_{\beta})=\left(
\begin{array}{cccc}
  0 &1&0&0 \\
  -1 &0&0&0 \\
  0&0&0&0\\
  0&0&0&0\\
 \end{array}\right)\hspace{0.1cm},\hspace{0.1cm}(Q_{\alpha\beta})=\left(
\begin{array}{cccc}
  0 &0&0&0 \\
  0 &0&0&0 \\
  0&0&0&-1\\
 0&0&1&0\\
 \end{array}\right),\hspace{0.1cm}(P^{\alpha\beta})=\left(
\begin{array}{cccc}
  0 &0&0&0 \\
  0 &0&0&0 \\
  0&0&0&-1\\
 0&0&1&0\\
 \end{array}\right).
\end{equation}
 Then the sigma model (\ref{gcsm1}) for this example, after omitting surface terms, is given by the following action:
\begin{equation*}
S=\int dzd\bar{z}[k_{0}\partial u\bar{\partial}u+\partial a_{1}\bar{\partial}a_{1}+\partial a_{2}\bar{\partial}a_{2}+\partial v\bar{\partial}u+\partial u\bar{\partial}v-\frac{a_{1}}{2}(\partial u\bar{\partial}a_{2}+\partial a_{2}\bar{\partial}u)+\frac{a_{2}}{2}(\partial u\bar{\partial}a_{1}+\partial a_{1}\bar{\partial}u)
\notag\end{equation*}
\begin{equation}\label{sa410}
-(k'-k'')u(\partial a_{1}\bar{\partial}a_{2}-\partial a_{2}\bar{\partial}a_{1})].
\end{equation}
For investigating the integrability of this model correspond to the integrability condition (\ref{me}), by utilizing the Christoffel components
\begin{align}
\Gamma^{1}\hspace{0cm}_{23}=\frac{1}{2},\hspace{0.5cm} \Gamma^{2}\hspace{0cm}_{13}=-\frac{1}{2},\hspace{0.5cm}\Gamma^{4}\hspace{0cm}_{13}=-\frac{a_{1}}{4},\hspace{0.5cm}\Gamma^{4}\hspace{0cm}_{23}=-\frac{a_{2}}{4},
\notag\end{align}
 torsion components
 \begin{align}
H^{1}\hspace{0cm}_{32}=\frac{1}{2}(k'-k''),\hspace{0.5cm}H^{2}\hspace{0cm}_{13}=\frac{1}{2}(k'-k'')
,\hspace{0.5cm}H^{4}\hspace{0cm}_{12}=-\frac{1}{2}(k'-k''),\notag\end{align}
\begin{equation}\label{to}
 H^{4}\hspace{0cm}_{13}=\frac{a_{1}}{4}(k'-k''),\hspace{0.5cm}H^{4}\hspace{0cm}_{23}=\frac{a_{2}}{4}(k'-k''),
 \end{equation}
 and the matrices $ \alpha_{\mu}$ and $\mu_{\mu}$ by the relations (\ref{am}), we obtain the following conditions
\begin{equation*}
\lambda_{1}=\frac{\lambda'_{1}}{\lambda'_{4}}\lambda_{4},\hspace{0.5cm}\lambda_{3}=\frac{1}{2(\lambda_{4}+\lambda'_{4})}(\lambda'_{4}(1-k''+k'+2(\lambda_{2}-\lambda_{4}))+2\lambda_{4}(1+\lambda_{2}-\lambda_{4}))
,\end{equation*}
\begin{equation}
\lambda'_{3}=\frac{1}{2\lambda_{4}(\lambda_{4}+\lambda'_{4})}((k'-k''+1-2\lambda_{4}){\lambda'_{4}}^{2}-2\lambda_{4}\lambda'_{4}(\lambda_{4}-\lambda'_{2}-1)+2\lambda'_{2}{\lambda_{4}}^{2}),
\end{equation}
where $\lambda_{2},\lambda_{4},\lambda'_{1},\lambda'_{2},\lambda'_{4}$ are arbitrary constants, and one of them can be considered as spectral parameter.
Now comparing our model with the $WZW$ model on $\mathbf{A_{4,10}}$ Lie group \cite{NW}
\begin{equation*}
  S_{WZW_{A_{4,10}}}=\frac{K}{4\pi}\int dzd\bar{z}[k_{0}\partial u\bar{\partial}u+\partial v\bar{\partial}u+\partial u\bar{\partial}v+\partial a_{1}\bar{\partial}a_{1}+\partial a_{2}\bar{\partial}a_{2}-\frac{a_{1}}{2}(\partial u\bar{\partial}a_{2}+\partial a_{2}\bar{\partial}u)+\frac{a_{2}}{2}(\partial u\bar{\partial}a_{1}+\partial a_{1}\bar{\partial}u)
\notag\end{equation*}
\begin{equation}
+\frac{1}{2}u(\partial a_{1}\bar{\partial} a_{2}-\partial  a_{2}\bar{\partial} a_{1})].
\end{equation}
As in the previous example, by assuming $k'-k''=-\frac{1}{2}$ and $K=4\pi$ one can rewrite the sigma model (\ref{sa410}) as the above $WZW$ model .i.e. the perturbed term is equal to the $WZ$ term. For the case that
$k'-k''\neq-\frac{1}{2}$ the model (\ref{sa410}) is an integrable perturbed $WZW$ model:
\begin{equation}
S=S_{WZW_{A_{4,10}}}-(k'-k''+\frac{1}{2})\int dzd\bar{z}[u(\partial a_{1}\bar{\partial} a_{2}-\partial  a_{2}\bar{\partial} a_{1})].
\end{equation}
One can demonstrate that the models (\ref{sa481}), (\ref{sa482}) and (\ref{sa410}) exhibit conformal invariance up to one loop. In both models of $\mathbf{A_{4,8}}$ examples (\ref{sa481}) and (\ref{sa482}), there is only one non-zero component of the $Ricci$ tensor, $Ric_{11}=-\frac{1}{2}$, and six components of torsion (\ref{to1}) and (\ref{to2}).
 By setting $k'= -k_{0}$, $k=1$ or $k'= k_{0}$, $k=-1$  and $\phi=\text{const}$, all the components of the $\beta$ functions (\ref{bf}) become zero.
For the sigma model on the Lie group $\mathbf{A_{4,10}}$ (\ref{sa410}), there is only one non-zero component for the $Ricci$ tensor, $Ric_{33}=\frac{1}{2}$, and five non-zero components of the torsion (\ref{to}), this model satisfies $\beta$ function equations (\ref{bf}) by setting $k'=k''\pm 1$ and $\phi=const$.\\

 \section{\bf Perturbed $WZW$ model with generalized complex structure}
 As a generalization of the perturbed chiral model (\ref{gcsm1}), we propose a perturbed $WZW$ model with a generalized complex structure through the following action:
 \begin{equation}\label{gcsmp}
 S=S_{WZW}+\frac{K}{4\pi}\int dzd\bar{z} e^{\alpha}\hspace{0cm}_{\mu}(kg_{\alpha\delta}J^{\alpha}\hspace{0cm}_{\beta}+k'Q_{\alpha\beta}+k''g_{\alpha\delta}P^{\delta\sigma}g_{\sigma\beta})e^{\beta}\hspace{0cm}_{\nu}\partial x^{\mu}\bar{\partial}x^{\nu},
 \end{equation}
where for simplicity we assume $K=4\pi$. The torsion of the above model is the sum of the following torsion of the $WZW$ action (\ref{wzwt}) (as follows) and the torsion of the perturbed section obtained from (\ref{Hgcm})
\begin{equation}
H_{\mu\nu\lambda}=\frac{1}{2}g_{\alpha\delta} f^{\delta}\hspace{0cm}_{\beta\gamma}e^{\alpha}\hspace{0cm}_{\mu}e^{\beta}\hspace{0cm}_{\nu}e^{\gamma}\hspace{0cm}_{\lambda}\hspace{1mm},
\end{equation}
both examples of ($\mathbf{e}$) can be considered as the integrable perturbed $WZW$ model with generalized complex structure if

\begin{equation}\label{a48gccwzw}
\lambda_{1}= \frac{(kk_{0}+k'-2\lambda_{3})\lambda'_{3}-2\lambda_{3}^{2}}{2k_{0}(\lambda_{3}+\lambda'_{3})},\hspace{0.2cm} \lambda_{4}=\frac{(k'-2\lambda_{3})\lambda'_{3}-2\lambda_{3}(\lambda_{3}-k_{0})}{2k_{0}(\lambda_{3}+\lambda'_{3})},\hspace{0.2cm} \lambda'_{1}=\frac{\lambda_{1}}{\lambda_{3}}\lambda'_{3},\hspace{0.2cm} \lambda'_{4}=\frac{\lambda_{4}}{\lambda_{3}}\lambda'_{3},
\end{equation}
where $\lambda_{2},\lambda_{4},\lambda'_{1},\lambda'_{2},\lambda'_{4}$ are arbitrary constants, and one of them can be considered as spectral parameter.
Also, when investigating the integrability of the perturbed $WZW$ model in example ($\mathbf{f}$), we obtain the following conditions:

\begin{equation*}
\lambda_{1}=\frac{\lambda'_{1}}{\lambda'_{4}}\lambda_{4},\hspace{0.5cm}\lambda_{3}=\frac{1}{2(\lambda_{4}+\lambda'_{4})}(\lambda'_{4}(-k''+k'+2(\lambda_{2}-\lambda_{4}))+2\lambda_{4}(1+\lambda_{2}-\lambda_{4}))
,\end{equation*}
\begin{equation}
\lambda'_{3}=\frac{1}{2\lambda_{4}(\lambda_{4}+\lambda'_{4})}((-k''+k'-2\lambda_{4}){\lambda'_{4}}^{2}-2\lambda_{4}\lambda'_{4}(\lambda_{4}-\lambda'_{2}-1)+2\lambda'_{2}{\lambda_{4}}^{2})
,\end{equation}
where $\lambda_{2},\lambda_{4},\lambda'_{1},\lambda'_{2},\lambda'_{3},\lambda'_{4}$ are arbitrary constants, and one of them can be considered as spectral parameter.


\subsection{\bf Generalized complex structure on metric Lie algebra}\label{sec6}

In order to compare our models (\ref{gcsm1}) and (\ref{gcsmp}) with Mohammedi's work in \cite{NMO}, we attempt to reformulate the relations (\ref{pp}-\ref{D}) in terms of operators. On a metric Lie algebra $L$, it is possible to reformulate the conditions of generalized complex structure (\ref{pp}-\ref{D}) in terms of operators that act on $L$.
We consider the operators $J$, $\mathcal{P}$, and $\mathcal{Q}: L \rightarrow L$, which act on the generators $T_{\alpha}$ of $L$ as follows:
\begin{equation}
{J}(T_{\alpha})={J}^{\beta}\hspace{0.0cm}_{\alpha}T_{\beta}\hspace{0.5cm},\hspace{0.5cm}\mathcal{P}(T_{\alpha})=\mathcal{P}_{\alpha}\hspace{0.0cm}^{\beta}T_{\beta}\hspace{0.5cm},\hspace{0.5cm}\mathcal{Q}(T_{\alpha})=\mathcal{Q}_{\alpha}\hspace{0.0cm}^{\beta}T_{\beta},
\end{equation}
such that we define
\begin{equation}
\mathcal{P}_{\alpha}\hspace{0.0cm}^{\beta}=g_{\alpha\gamma}P^{\gamma\beta}\hspace{0.1cm},\hspace{0.1cm}\mathcal{Q}_{\alpha}\hspace{0.0cm}^{\beta}=Q_{\alpha\gamma}g^{\gamma\beta}
.\end{equation}
Here, $g_{\alpha\beta} = <T_{\alpha}, T_{\beta}>$ represents an ad-invariant invertible metric on the Lie algebra $L$. In this way, after some calculation, one can rewrite the conditions (\ref{pp}-\ref{D}) for a generalized complex structure on a metric Lie algebra $L$ as the following operator relations on $L$:
\begin{equation}\label{gco1}
\mathcal{P}^{t}=-\mathcal{P}\hspace{1cm},\hspace{1cm}\mathcal{Q}^{t}=-\mathcal{Q},
\end{equation}
\begin{equation}
{J}^{2}+\mathcal{P}\mathcal{Q}+I=0,
\end{equation}
\begin{equation}
({J\mathcal{P}})^{t}=-({J\mathcal{P}}),
\end{equation}
\begin{equation}
({J\mathcal{Q}})^{t}=-({J\mathcal{Q}}),
\end{equation}
$\forall X,Y \in L:$
\begin{equation}
[\mathcal{P}(X),\mathcal{P}(Y)]-\mathcal{P}[X,\mathcal{P}(Y)]-\mathcal{P}[\mathcal{P}(X),Y]=0,
\end{equation}
\begin{equation}
[\mathcal{P}(X),J(Y)]+[J(X),\mathcal{P}(Y)]=-J^t([\mathcal{P}(X),Y]+[X,\mathcal{P}(Y)]),
\end{equation}
\begin{equation}
[X,Y]-[J(X),J(Y)]+J[J(X),Y]+J[X,J(Y)]+\mathcal{P}[X,\mathcal{Q}(Y)]+\mathcal{P}[\mathcal{Q}(X),Y]=0,
\end{equation}
\begin{equation}\label{gco7}
[J(X),\mathcal{Q}(Y)]+[\mathcal{Q}(X),J(Y)]+J^t([\mathcal{Q}(X),Y]+[X,\mathcal{Q}(Y)])-\mathcal{Q}([J(X),Y]+[X,J(Y)])=0,
\end{equation}
where transpose of an operator $O$ is defined as
\begin{equation}
 \forall X,Y \in L \hspace{0.1cm},\hspace{0.5cm}<OX,Y>=<X,O^tY>.
\end{equation}
Furthermore one can rewrite our sigma model (\ref{gcsm1}) as the following form:
\begin{equation}\label{gcmm}
S=\int dzd\bar{z}[<l^{-1}\partial l,l^{-1} \bar{\partial} l>+<l^{-1}\partial l,(kJ^t-k'\mathcal{Q}-k''\mathcal{P})l^{-1} \bar{\partial} l>],
\end{equation}
in this form, one can compare our model with the model presented in \cite{NMO}. When comparing (\ref{gcmm}) with the model (22) in \cite{NMO} (with $\lambda=0$), we find that $Q'^{-1}=I+kJ^t-k'\mathcal{Q}-k''\mathcal{P}$ and $P'=Q'^{t}$ in (22) of \cite{NMO}\footnote{To prevent confusion, we use the operators $P'$ and $Q'$ here instead of $P$ and $Q$ as used in \cite{NMO}.}. Here, we consider result of the first example $(\mathbf{e})$ as follows:
\begin{equation}
Q'^{-1}=I+kJ^t-k'\mathcal{Q}-k''\mathcal{P}=\left(
\begin{array}{cccc}
  1+\frac{k'}{k_{0}} &0&-k+\frac{k'}{k_{0}}&0 \\
  0 &1+\frac{k'}{k_{0}}&0&-k+\frac{k'}{k_{0}} \\
  k+\frac{k'}{k_{0}}&0&1-\frac{k'}{k_{0}}&0\\
  0&k+\frac{k'}{k_{0}}&0&1-\frac{k'}{k_{0}}\\
 \end{array}\right).
\end{equation}
One can check that the above operators $Q'$ and $P'=Q'^t$ satisfy the relation (21) in \cite{NMO}.
In this manner, our model, which is only on a metric Lie algebra, is equivalent to \cite{NMO}. The operators $J$, $\mathcal{P}, \mathcal{Q}$ must satisfy the relations of generalized complex structure (\ref{gco1}-\ref{gco7}). To investigate the integrability condition (\ref{me}) we use the $\alpha_{\mu} $ and $ \mu_{\mu}$ matrices
by the relations (\ref{am}), so the same conditions obtain which obtain for the model (\ref{sa481})\footnote{Of course with negative sign for the $k$ and $k'$ in the action (\ref{gcsm1}).}.
One can also perform this work for the perturbed $WZW$ model (\ref{gcsmp}). Note that one can construct other integrable sigma models by different choices of $Q'$ which some of them are under investigation in other work.
\section{\bf Conclusion}\label{sec7}
Using the general method presented by Mohammedi \cite{NM} for the integrability of a sigma model on a manifold, we obtain conditions for having an integrable deformation of a general sigma model on a manifold with a complex structure. We show that on a Lie group, these conditions are satisfied by using the zeros of the Nijenhuis tensor. We then extend this formalism to models on both manifolds and Lie groups with a generalized complex structure. We compare our results with those presented by Mohammedi \cite{NMO}. Similar investigations can be carried out with Poisson structures and Poisson-Nijenhuis structures \cite{KE} on a Lie group \cite{RRH}, allowing the construction of integrable deformations of sigma models with these structures. These investigations are currently underway.
\section{\bf Appendix}
Here, we will prove relation (\ref{C}) by applying relations (39) and (86) to equation (\ref{njg3}). Thus, we have:
\begin{equation*}
e_{\delta}\hspace{0cm}^{\lambda}J^{\delta}\hspace{0cm}_{\sigma}e^{\sigma}\hspace{0cm}_{\nu}\partial_{\lambda}(e_{\delta'}\hspace{0cm}^{\mu})J^{\alpha'}\hspace{0cm}_{\sigma'}e^{\sigma'}\hspace{0cm}_{k}+e_{\delta}\hspace{0cm}^{\lambda}J^{\delta}\hspace{0cm}_{\sigma}e^{\sigma}\hspace{0cm}_{\nu}e_{\alpha'}\hspace{0cm}^{\mu}J^{\delta'}\hspace{0cm}_{\sigma'}\partial_{\lambda}(e^{\sigma'}\hspace{0cm}_{k})
\end{equation*}
\begin{equation*}
-e_{\delta}\hspace{0cm}^{\lambda}J^{\delta}\hspace{0cm}_{\sigma}e^{\sigma}\hspace{0cm}_{k}\partial_{\lambda}(e_{\delta'}\hspace{0cm}^{\mu})J^{\delta'}\hspace{0cm}_{\sigma'}e^{\sigma'}\hspace{0cm}_{\nu}-e_{\delta}\hspace{0cm}^{\lambda}J^{\delta}\hspace{0cm}_{\sigma}e^{\sigma}\hspace{0cm}_{k}e_{\delta'}\hspace{0cm}^{\mu}J^{\delta'}\hspace{0cm}_{\sigma'}\partial_{\lambda}(e^{\sigma'}\hspace{0cm}_{\nu})
\end{equation*}
\begin{equation*}
-e_{\alpha}\hspace{0cm}^{\mu}J^{\delta}\hspace{0cm}_{\sigma}e^{\sigma}\hspace{0cm}_{\lambda}\partial_{\nu}(e_{\delta'}\hspace{0cm}^{\lambda})J^{\delta'}\hspace{0cm}_{\sigma'}e^{\sigma'}\hspace{0cm}_{k}-e_{\delta}\hspace{0cm}^{\mu}J^{\delta}\hspace{0cm}_{\sigma}e^{\beta}\hspace{0cm}_{\lambda}e_{\delta'}\hspace{0cm}^{\lambda}J^{\delta'}\hspace{0cm}_{\sigma'}\partial_{\nu}(e^{\sigma'}\hspace{0cm}_{k})
\end{equation*}
\begin{equation*}
+e_{\delta}\hspace{0cm}^{\mu}J^{\delta}\hspace{0cm}_{\sigma}e^{\sigma}\hspace{0cm}_{\lambda}\partial_{k}(e_{\delta'}\hspace{0cm}^{\lambda})J^{\delta'}\hspace{0cm}_{\sigma'}e^{\sigma'}\hspace{0cm}_{\nu}+e_{\delta}\hspace{0cm}^{\mu}J^{\delta}\hspace{0cm}_{\sigma}e^{\sigma}\hspace{0cm}_{\lambda}e_{\delta'}\hspace{0cm}^{\lambda}J^{\delta'}\hspace{0cm}_{\sigma'}\partial_{k}(e^{\sigma'}\hspace{0cm}_{\nu})
\end{equation*}
\begin{equation*}
+e_{\delta}\hspace{0cm}^{\mu}P^{\delta\sigma}e_{\sigma}\hspace{0cm}^{\lambda}\partial_{\lambda}(e^{\delta'}\hspace{0cm}_{\nu})Q_{\delta'\sigma'}e^{\sigma'}\hspace{0cm}_{k}+e_{\delta}\hspace{0cm}^{\mu}P^{\delta\sigma}e_{\sigma}\hspace{0cm}^{\lambda}e^{\delta'}\hspace{0cm}_{\nu}Q_{\delta'\sigma'}\partial_{\lambda}(e^{\sigma'}\hspace{0cm}_{k})
\end{equation*}
\begin{equation*}
+e_{\delta}\hspace{0cm}^{\mu}P^{\delta\sigma}e_{\sigma}\hspace{0cm}^{\lambda}\partial_{\nu}(e^{\delta'}\hspace{0cm}_{k})Q_{\delta'\sigma'}e^{\sigma'}\hspace{0cm}_{\lambda}+e_{\delta}\hspace{0cm}^{\mu}P^{\delta\sigma}e_{\sigma}\hspace{0cm}^{\lambda}e^{\delta'}\hspace{0cm}_{k}Q_{\delta'\sigma'}\partial_{\nu}(e^{\sigma'}\hspace{0cm}_{\lambda})
\end{equation*}
\begin{equation*}
+e_{\delta}\hspace{0cm}^{\mu}P^{\delta\sigma}e_{\sigma}\hspace{0cm}^{\lambda}\partial_{k}(e^{\delta'}\hspace{0cm}_{\lambda})Q_{\delta'\sigma'}e^{\sigma'}\hspace{0cm}_{\nu}+e_{\delta}\hspace{0cm}^{\mu}P^{\delta\sigma}e_{\sigma}\hspace{0cm}^{\lambda}e^{\delta'}\hspace{0cm}_{\lambda}Q_{\delta'\sigma'}\partial_{k}(e^{\sigma'}\hspace{0cm}_{\nu})=0.
\end{equation*}
Now, we multiply the above relation by $e_{\alpha}\hspace{0.0cm}^{\nu}e_{\beta}\hspace{0.0cm}^{k}e^{\gamma}\hspace{0cm}_{\mu}$ and sum over the indices $\mu$, $\nu$ and $k$ and using that $e^{\delta}\hspace{0.0cm}_{\mu}e_{\sigma}\hspace{0.0cm}^{\mu}=\delta^{\delta}\hspace{0.0cm}_{\sigma}\hspace{0.1cm}, \hspace{0.1cm} e^{\delta}\hspace{0.0cm}_{\mu}e_{\delta}\hspace{0.0cm}^{\nu}=\delta_{\mu}\hspace{0.0cm}^{\nu}
$ and using (\ref{jj}) in second terms of the third and forth lines of above relation, we obtain the relation in the following form
\begin{equation*}
e_{\delta}\hspace{0cm}^{\lambda}J^{\delta}\hspace{0cm}_{\alpha}
\partial_{\lambda}(e_{\sigma}\hspace{0cm}^{\mu})J^{\sigma}\hspace{0cm}_{\beta}e^{\gamma}\hspace{0cm}_{\mu}
+e_{\delta}\hspace{0cm}^{\lambda}J^{\delta}\hspace{0cm}_{\alpha}J^{\gamma}\hspace{0cm}_{\sigma}\partial_{\lambda}(e^{\sigma}\hspace{0cm}_{k})e_{\beta}\hspace{0cm}^{k}
\end{equation*}
\begin{equation*}
-e_{\delta}\hspace{0cm}^{\lambda}J^{\delta}\hspace{0cm}_{\beta}\partial_{\lambda}(e_{\sigma}\hspace{0cm}^{\mu})J^{\sigma}\hspace{0cm}_{\alpha}e^{\gamma}\hspace{0cm}_{\mu}
-e_{\delta}\hspace{0cm}^{\lambda}J^{\delta}\hspace{0cm}_{\beta}J^{\gamma}\hspace{0cm}_{\sigma}\partial_{\lambda}(e^{\sigma}\hspace{0cm}_{\nu})e_{\alpha}\hspace{0cm}^{\nu}
\end{equation*}
\begin{equation*}
-J^{\gamma}\hspace{0cm}_{\sigma}e^{\sigma}\hspace{0cm}_{\lambda}\partial_{\nu}(e_{\delta}\hspace{0cm}^{\lambda})J^{\delta}\hspace{0cm}_{\beta}e_{\alpha}\hspace{0cm}^{\nu}+e_{\alpha}\hspace{0cm}^{\nu}e_{\beta}\hspace{0cm}^{k}\partial_{\nu}(e^{\gamma}\hspace{0cm}_{k})+P^{\gamma\delta}Q_{\delta\sigma}\partial_{\nu}(e^{\sigma}\hspace{0cm}_{k})e_{\alpha}\hspace{0.0cm}^{\nu}e_{\beta}\hspace{0cm}^{k}
\end{equation*}
\begin{equation*}
+J^{\gamma}\hspace{0cm}_{\sigma}e^{\sigma}\hspace{0cm}_{\lambda}\partial_{k}(e_{\delta}\hspace{0cm}^{\lambda})J^{\delta}\hspace{0cm}_{\alpha}e_{\beta}\hspace{0cm}^{k}-e_{\alpha}\hspace{0cm}^{\nu}e_{\beta}\hspace{0cm}^{k}\partial_{k}(e^{\gamma}\hspace{0cm}_{\nu})-P^{\gamma\delta}Q_{\delta\sigma}\partial_{k}(e^{\sigma}\hspace{0cm}_{\nu})e_{\alpha}\hspace{0.0cm}^{\nu}e_{\beta}\hspace{0cm}^{k}
\end{equation*}
\begin{equation*}
+P^{\gamma\sigma}e_{\sigma}\hspace{0cm}^{\lambda}\partial_{\lambda}(e^{\delta}\hspace{0cm}_{\nu})Q_{\delta\beta}e_{\alpha}\hspace{0.0cm}^{\nu}+P^{\gamma\sigma}e_{\sigma}\hspace{0cm}^{\lambda}\partial_{\lambda}(e^{\delta}\hspace{0cm}_{k})Q_{\alpha\delta}e_{\beta}\hspace{0.0cm}^{k}
\end{equation*}
\begin{equation*}
+P^{\gamma\sigma}\partial_{\nu}(e^{\delta}\hspace{0cm}_{k})Q_{\delta\sigma}e_{\alpha}\hspace{0.0cm}^{\nu}e_{\beta}\hspace{0.0cm}^{k}+P^{\gamma\sigma}e_{\sigma}\hspace{0cm}^{\lambda}\partial_{\nu}(e^{\delta}\hspace{0cm}_{\lambda})Q_{\beta\delta}e_{\alpha}\hspace{0.0cm}^{\nu}
\end{equation*}
\begin{equation*}
+P^{\gamma\sigma}e_{\sigma}\hspace{0.0cm}^{\lambda}\partial_{k}(e^{\delta}\hspace{0cm}_{\lambda})Q_{\delta\alpha}e_{\beta}\hspace{0.0cm}^{k}+P^{\gamma\sigma}Q_{\sigma\delta}\partial_{k}(e^{\delta}\hspace{0cm}_{\nu})e_{\alpha}\hspace{0cm}^{\nu}e_{\beta}\hspace{0.0cm}^{k}=0,
\end{equation*}
then by noting that $P^{\alpha\beta}$ and $Q_{\alpha\beta}$ are antisymmetric and using the identities $\partial_{k}(e_{\alpha}\hspace{0.0cm}^{\lambda})e^{\beta}\hspace{0.0cm}_{\lambda}+e_{\alpha}\hspace{0.0cm}^{\lambda}\partial_{k}(e^{\beta}\hspace{0.0cm}_{\lambda})=0, e^{\delta}\hspace{0.0cm}_{\mu}e_{\sigma}\hspace{0.0cm}^{\mu}=\delta^{\delta}\hspace{0.0cm}_{\sigma}$ along with the Maurer–Cartan equation (\ref{mc}), one can, after some calculations, obtain relation (\ref{C}).

\subsection*{Acknowledgements}

This work has been supported by the research vice chancellor of Azarbaijan Shahid Madani University under research fund No. 1402/231.

\end{document}